\documentclass[fleqn,usenatbib]{mnras}
\usepackage{newtxtext,newtxmath}
\usepackage[T1]{fontenc}
\usepackage{soul}
\usepackage{comment}
\usepackage{multirow}
\DeclareRobustCommand{\VAN}[3]{#2}
\let\VANthebibliography\thebibliography
\def\thebibliography{\DeclareRobustCommand{\VAN}[3]{##3}\VANthebibliography}

\newcommand{\teff}{$\mathrm{T_{eff}}$}
\newcommand{\logg}{$\log$~g}
\newcommand{\vsini}{$v \sin i$}

\newcommand{\tona}{{\tt TONALLI}}
\newcommand{\alfa}{[$\alpha$/M]}
\newcommand{\meta}{[M/H]}
\newcommand{\bacc}{{\tt BACCHUS}}
\usepackage{lineno}
\usepackage{graphicx}	
\usepackage{amsmath}
\usepackage{natbib}
\defcitealias{lopezval25}{Paper~I}

\title[Parameters and abundances of YSOs II. C, N, O]{Atmospheric parameters and chemical abundances of young stars with APOGEE -- II. Carbon, Nitrogen, Oxygen, and Sulphur abundances in Orion}
\author[L\'opez-Valdivia et al.]{
Ricardo L\'opez-Valdivia,$^{1}$
A. Z. Lugo-Aranda,$^{1,2}$
Sergio S\'anchez-Sanju\'an,$^{1}$
J. E. M\'endez-Delgado,$^{3}$
\newauthor{Carlos G. Rom\'an-Z\'u\~niga,$^{1}$}\thanks{E-mail: croman@astro.unam.mx}
Luc\'ia Adame,$^{1}$
Jes\'us Hern\'andez,$^{1}$
Edilberto S\'anchez,$^{1}$,
\newauthor{Itzarel Hern\'andez-Aburto,$^{1}$}
{Jos\'e G. Fern\'andez-Trincado,$^{4}$}
Leticia Carigi,$^{3}$
Richard R. Lane,$^{4}$
Marina Kounkel,$^{5}$
\newauthor{Javier Serna,$^{6}$}
Keivan G. Stassun,$^{7}$
{Sandro Villanova,$^{2}$}
{Jinyoung Serena Kim,$^{8}$}
Jonathan C. Tan$^{9,10}$
\\
$^{1}$Universidad Nacional Autónoma de México. Instituto de Astronomía. A.P. 106, 22800. Ensenada, B.C. , México\\
$^{2}$Universidad Andres Bello, Facultad de Ciencias Exactas, Departamento de F{\'i}sica y Astronom{\'i}a - Instituto de Astrof{\'i}sica, \\ Autopista Concepci\'on-Talcahuano 7100, Talcahuano, Chile\\
$^{3}$Universidad Nacional Autónoma de México. Instituto de Astronomía. A.P. 70-264, 04510. Ciudad de México, México\\
$^{4}$Centro de investigaci\'on en Astronom\'ia, Facultad de Ingenier\'ia, Ciencia y Tecnolog\'ia, Universidad Bernardo O’Higgins, Av. Viel 1497, Santiago, 8370993, Chile \\
$^5$Department of Physics and Astronomy, University of North Florida, 1 UNF Dr, Jacksonville, FL 32224, USA \\
$^6$Homer L. Dodge Department of Physics and Astronomy, University of Oklahoma, Norman, OK 73019, USA\\
$^7$Department of Physics and Astronomy, Vanderbilt University, VU Station 1807, Nashville, TN 37235, USA\\
$^{8}$Steward Observatory, University of Arizona, 933 N. Cherry Ave., Tucson,
AZ 85721-0065, USA \\
$^{9}$Dept. of Astronomy, Univ. of Virginia, Charlottesville, VA 2904, USA\\
$^{10}$Dept. of Physics \& Astronomy, Chalmers Univ. of Technology, E-412 96 Gothenburg, Sweden\\
}

\date{Accepted XXX. Received YYY; in original form ZZZ}

\pubyear{\the\year{}}

\begin{document}
\label{firstpage}
\pagerange{\pageref{firstpage}--\pageref{lastpage}}
\maketitle

\begin{abstract}
We derived the abundances of carbon, nitrogen, oxygen, sulphur, and iron in young G-, K-, and M-type stars (effective temperatures of 3500--5900 K) using APOGEE-2 infrared spectra. Our final sample comprises 195 slow rotators (\vsini\ $\leq$ 30 km s$^{-1}$) in the Orion complex (Orion A, B, OB1, and $\lambda$~Ori). The abundances were determined using MARCS model atmospheres, atmospheric parameters derived with \tona, and the \bacc\ code. {We obtained median abundances of $\epsilon$(C) = 8.37$\pm$0.10, $\epsilon$(N) = 7.35$\pm$0.16, $\epsilon$(O) = 8.63$\pm$0.08, $\epsilon$(S) = 7.19$\pm$0.18, and $\epsilon$(Fe) = 7.45$\pm$0.05.} Nitrogen {is} subsolar {([N/H] $=-0.48$)}, whereas carbon, oxygen, sulphur, and iron are {within 0.1 dex of the solar values}. We find no spatial structure within the Orion subgroups, and the only significant difference is a {0.29 dex} lower nitrogen abundance in Orion A {than in Orion OB1}, which does not extend to the other elements. Compared with {430} nearby main-sequence stars analysed in the same way, Orion shows {carbon, nitrogen, and oxygen abundances lower by 0.1--0.2 dex}, although consistent within the errors, whereas sulphur and iron {differ by less than 0.07 dex}. We also compare our abundances with nebular abundances of HII regions and star-forming galaxies: the stellar abundances of oxygen, sulphur, and iron are higher, with the largest offset in iron, although the agreement in abundance ratios improves when temperature inhomogeneities and dust depletion are taken into account.

\end{abstract}




\section{Introduction}
The Orion star-forming region, at a distance of 400 pc, has long served as a benchmark for testing chemical homogeneity in young stellar populations \citep[e.g. ][]{blaauw64,brown95,stassun99,dario16,dario17,kounkel18,hernandez23}. Early studies have reported mixed results, ranging from element-specific abundance variations \citep{cunha94,cunha98} to abundances consistent with solar or thin-disk values \citep{simon10,dorazi09,biazo11,biazo12}. However, the conclusions drawn in those early studies could have been affected by small-number statistics, as available samples in years prior to large-scale spectroscopic surveys, were limited to only a few dozen objects.

Over the last $\sim$15 years, the astronomical community has had access to large spectroscopic surveys such as LAMOST \citep{cui12,zhao12}, Gaia-ESO \citep{gilmore12}, GALAH \citep{silva15,martell17}, and the different phases of the Sloan Digital Sky Survey \citep{blanton17,majewski17,almeida23,kollmeier17}, which have increased the number of observed young stellar objects, allowing for detailed chemical analyses in star-forming regions based on hundreds of objects.

For instance, \citet{kos21} derived atmospheric parameters and chemical abundances of 25 elements in a GALAH sample of $\sim$300 stars. They found no evidence of self-enrichment across different Orion subgroups, although significant trends of abundances with $T_{\rm eff}$ were reported, possibly reflecting unaccounted systematic effects. 

The SDSS-IV APOGEE-2 survey operates in the $H$-band ($\sim$1.5–1.7 $\mu$m), making it particularly well suited for the analysis of recently formed stars in nearby star-forming regions \citep[e.g. ][]{foster15,cottar15,fernandez17,roman19,roman-zuniga23,lopezval25}. Young and pre-main-sequence stars (ages $\lesssim$ 300 Myr) are affected by a number of physical processes, such as magnetic activity \citep[e.g.  ][]{sokal18,lopezval23}, high rotation \citep[e.g.  ][]{King00,nofi21}, and variability \citep[e.g.  ][]{cody22}, which together complicate the derivation of atmospheric parameters and reliable chemical abundances. Near-infrared spectroscopy, especially in the $H$-band region provides an important advantage for the study of young stellar populations, as {SDSS-APOGEE} stellar spectra are significantly less affected by extinction and continuum veiling than optical observations (Zepeda-Villanueva et al., in prep).

In a previous study, \citet[][herafter Paper I]{lopezval25} determined the atmospheric parameters for 548 young G-, K-, and M-type stars located in four Orion subgroups (Orion A, B, OB1, and $\lambda$~Ori). For a subset of 340 slow rotators (\vsini\ $\leq 30$ km s$^{-1}$), we were able to derive C, Mg, Si, K, Ti, and Fe abundances. We found slightly subsolar [X/H] abundance ratios (median values between -0.22 and -0.03 dex), consistent across elements and among all four groups. Furthermore, the median stellar [C/H] abundance derived for Orion agrees with previous estimates based on analyses of the ionized gas in the Orion Nebula. Although carbon might not be the most common element to use for comparison between stellar and ionized gas abundances, our previous results suggest that both quantities are consistent.

Among all the chemical elements, nitrogen, oxygen, and sulphur are the most commonly measured elements in H {\sc ii} regions \citep[e.g. ][]{floridozuritaperez2022,gavilandiazzamora2025,sattlermendezkreckel2026} and star forming galaxies \citep[e.g. ][]{diazzamora2022,arellanocullencarnall2025}, as they have some of the strongest nebular emission lines, and are widely used in abundance studies of photoionized environments. 

Comparisons between stellar and nebular abundances in Orion and similar nearby regions suggest that, within uncertainties, the photospheric abundances of young stars might be consistent with those measured in the surrounding ionized gas, especially for oxygen and nitrogen \citep[e.g. ][]{cunha93,garcia-rojas14}. Nevertheless, this kind of comparison is often made with the massive stars that ionize the H {\sc ii} regions; to date, there is no {dedicated} study exploring the chemical abundances of low-mass stars and how these stars relate to the surrounding ionized gas. In addition, as mentioned above, young stars are affected by several physical processes, while ionized-gas analyses are subject to uncertainties such as temperature fluctuations \citep{Peimbert:1967,MartinezHernandez:2026} and abundance discrepancy factors \citep{Liu00,GarciaRojasEsteban07,MendezDelgado23}, making direct comparisons between stellar and gas-phase abundances particularly challenging.

In this work, we extend our previous stellar abundance analysis of Orion to include N, O, and S. We aim to directly link the photospheric compositions of low-mass young stars with the chemical properties of the ionized gas to provide a more comprehensive view of elemental enrichment in both components.

{The paper is organized as follows. In Section \ref{sect:sample}, we describe the stellar sample used in this study, while in Section \ref{sect:method}, we present the details of our chemical abundance determination method and discuss our results. In Section \ref{sect:compar}, we compare our chemical abundances with those of stars in the solar neighborhood and with nebular determinations. Finally, in Section \ref{sect:summary} {we summarize our results and present our main conclusions in Section \ref{sect:conclusions}.}

\section{Sample}\label{sect:sample}
{Our working sample is the same photospheric, infrared excess-free sample constructed in \citetalias{lopezval25}, to which we refer the reader for a complete description; we summarize it here for completeness.} 

{The sample originates from the APOGEE-2 survey \citep{majewski17}, from which \cite{roman-zuniga23} constructed a catalog of more than 3000 young stars spanning spectral types from {G to M} and distributed across 16 Galactic star-forming regions and young open clusters. In the case of Orion, \cite{roman-zuniga23} found, through a clustering-based selection method, a robust sample of 1861 bona fide members.} 

{Young stars frequently show infrared excess from interstellar extinction or a protostellar disk, both linked to processes (e.g.  accretion, magnetic activity, and continuum veiling) that can bias the derived atmospheric parameters and thus chemical abundances. Veiling is non-stellar emission that fills in photospheric absorption lines \citep{joy49}; in the infrared it arises from dust emission at the inner disk's sublimation radius \citep[e.g. ][]{natta01,muzerolle03} and from warm gas interior to it \citep[e.g. ][]{fischer11}. To minimize the impact of veiling in our analysis, we applied a semi-empirical photometric selection. We used 2MASS \citep{2mass} and WISE \citep{wise} photometry, as compiled in the \texttt{ALLWISE} catalog \citep{cutri13}, to build a $J$ vs. $(K-W3)$ color-magnitude diagram and identify stars that depart from the expected, veiling-free photospheric color derived from a main-sequence star sample. This photospheric color is approximately constant, $(K-W3)_{\rm phot} \simeq 0.267$~mag, for $J<8$~mag, but for fainter objects it follows the fourth-degree polynomial}
\begin{equation}
(K-W3)_{\rm phot} = a + bJ + cJ^2 + dJ^3 + eJ^4
\label{eq:poli_veiling}
\end{equation}
with $a=0.5138$, $b=-0.2258$, $c=0.0713$, $d=-0.0091$, and $e=0.0004$.

{In \citetalias{lopezval25} we kept only sources with reliable, uncontaminated WISE photometry (quality flag A, B, or C; ccf,$=$,0) whose $(K-W3)$ color agreed with the photospheric value within 1.5 times its uncertainty, yielding a sample of 559 stars with negligible infrared excess.} Finally, we retrieved the epoch-combined, one-dimensional infrared spectra of these objects from the APOGEE-2 DR17 public archive. 

{We assessed the possible impact of ongoing accretion, another physical process that could bias our determinations. We inspected the $H$-band spectra for emission from the hydrogen Brackett 11 (Br11) line ($\lambda\sim$16811 \AA). For cool young stars, detectable Br11 emission can provide evidence of hot circumstellar gas associated with accretion. 

{We found no significant Br11 emission in the spectra of the selected stars. {{Br11 emission is considered detectable in APOGEE spectra when its equivalent width exceeds $\sim$0.75~\AA\ \citep{campbell23}. Converting this equivalent width into a line luminosity, and then into an accretion luminosity with the $L_{\rm acc}$--$L_{\rm Br11}$ relation of \citet{fiorellino25}, we estimate that our non-detections correspond to mass accretion rates below $\dot{M}_{\rm acc}\sim10^{-8.5}$~M$_\odot$~yr$^{-1}$ for stellar masses between 0.5 and 1~M$_\odot$.}Therefore, we do not expect it to bias our stellar atmospheric parameters and chemical abundances, although this does not rule out accretion at a level below our detection threshold.}

The basic atmospheric parameters (namely effective temperature \teff, surface gravity \logg, overall metallicity \meta, $\alpha$-elements enhancement \alfa, and projected rotational velocity \vsini) were derived using the \tona\ code \citep{adame24}. \tona\ implements the optimization algorithm (the asexual genetic algorithm) of \cite{canto09} to fit synthetic spectra to observations through a $\chi^2$ minimization. The code interpolates synthetic spectra from the grid of \citet{jonsson20}, applies rotational broadening following \citet{gray08}, and iteratively optimizes the parameters until convergence is achieved. 
In order to obtain better \logg\ estimates with \tona, in \citetalias{lopezval25} we used photometric estimates obtained with the \texttt{TEPITZIN} module (Adame et al., in prep.), which compares observed Gaia and 2MASS absolute magnitudes with the derivatives of the PARSEC evolutionary models \citep{PARSEC,PARSEC22}. 
The photometry was dereddened by estimating the interstellar extinction for the full sample using the \texttt{MASSAGE} code \citep{hernandez23,massage26} 

The final atmospheric parameters were derived after {30} \tona\ runs, adopting the medians and inter-quartile ranges of the posterior distributions as best values and uncertainties. 
Out of the initial 559 stars analyzed, 548 yielded reliable results. These correspond to stars with inter-quartile ranges in \teff\ and \vsini\ smaller than 500 K and 5 km s$^{-1}$, respectively.
The derived effective temperatures range from 3200 to 6000 K, with surface gravities consistent with the expected values for young stellar objects in Orion.

As in \citetalias{lopezval25}, we selected those stars with \vsini\ $\leq$ 30 km s$^{-1}$, as this velocity threshold ensures that the stars will not show significant line blending in their spectra, thus increasing the reliability of our chemical abundance estimates. Therefore, our working sample has 361 stars, distributed as follows: 156 stars located in the Orion~OB1 group, 82 stars in $\lambda$~Ori, 68 stars in Orion~A, and 55 stars in Orion~B (see Figure 1 of \citetalias{lopezval25}). 

{Our four Orion groups span a range of ages between 3 and 11 Myr (see Section \ref{sect:age}), which allows us to search for abundance differences as a function of location and age among the Orion complex; if present, such differences could be closely tied to evolutionary and chemical enrichment processes, acting on the relatively short timescales.}

\section{Chemical Abundance Analysis}\label{sect:method}
To determine the chemical abundances of our Orion sample, we adopted the atmospheric parameters derived in \citetalias{lopezval25} and the reduced spectra of APOGEE DR17 in the Brussels automated code to characterize high-accuracy spectra \citep[{\tt BACCHUS};][]{masseron16}.

\bacc\ computes on-the-fly synthetic spectra for a range of abundances that are compared to the observed spectrum on a line-by-line basis. The version of \bacc\ we used relies on the v15.1 of the 1D local thermodynamic equilibrium (LTE) radiative transfer code {\tt TURBOSPECTRUM} \citep{turbo}. To generate the synthetic spectra we employed the MARCS model atmosphere grid \citep{jonsson20}, the atomic line list of \cite{smith21}, and the molecular line lists for CO, CN, and OH from \cite{li15}, \cite{sneden14}, and \cite{brooke16}, respectively. 

To compute a synthetic spectrum, \texttt{TURBOSPECTRUM} requires the stellar parameters \teff, \logg, \meta, the microturbulent velocity, a convolution parameter (instrumental + rotational broadening), and an adopted solar composition, which in our case follows \citet*{grevese07}. As in \citetalias{lopezval25}, we adopted the {\tt BACCHUS} method that minimizes the squared residuals between observed and synthetic spectra as it provides the most balanced results among the four methods available within the code. 

For each star in our sample, we used the epoch-combined 1D APOGEE spectrum\footnote{The \texttt{apStar} spectrum files are available at the SDSS-IV science archive web app (SAW)},  the \tona\ atmospheric results \teff, \logg, \meta, and {a line list conformed by atomic and molecular transitions to determine the abundance of C, N, O, S, and Fe.} 
We treated the microturbulence velocity and the convolution parameter of \bacc\ as free parameters; in order to increase the convergence effectiveness of \bacc, we initialized the convolution parameter to be equal to the \vsini\ value found by \tona.

{To determine the chemical abundances, we employed an iterative procedure designed to preserve the molecular equilibrium of CO, CN, and OH, which are molecules that couple the abundances of carbon, nitrogen, and oxygen. We started by determining the iron abundance from a set of Fe lines. The resulting iron abundance was then fixed and used to derive the oxygen abundance from OH molecular features. Subsequently, the oxygen abundance was held fixed while the carbon abundance was determined from CO lines. The derived carbon abundance was then fixed together with oxygen, and the nitrogen abundance was obtained from CN transitions. Finally, the iron abundance was recomputed and compared with the value adopted at the beginning of the iteration. The entire sequence was repeated until the difference between two consecutive iron abundance determinations became smaller than 1\% of the initial value} {or the maximum permitted number of iterations was reached, which in our case was 15, a number that offers a good compromise between computational cost and convergence rate, as most stars reached convergence by iteration 5 or earlier. Finally, sulphur abundances were independently derived from selected S lines. We considered a run converged if it had simultaneous C, N, O, and Fe abundance determinations.}

{We first applied the mentioned above iterative procedure to the complete sample using the complete set of available spectral features. This initial run was used to identify the most reliable lines for each element.} In particular, we identified spectral features that did not exhibit strong (Spearman coefficient $r_{\rm S}\leq$ 0.4) trends between chemical abundances and the stellar parameters \teff\ and \logg\ \citep[e.g.  ][]{kos21,vinicius24,kos25}. We relaxed this criterion for sulphur to $r_{\rm S}\leq 0.45$, as most of the S lines showed higher correlation coefficients. The basic information of the final atomic and molecular line list considered in this work is detailed in Table \ref{tab:lines}.

{Once this line selection was established, we repeated the abundance analysis using only the selected features, this time including a Monte Carlo procedure to propagate the uncertainties associated with the atmospheric parameters.} For each star, we repeated the abundance iterative analysis one hundred times, randomly perturbing the atmospheric parameters (\teff, \logg, \meta, and \vsini) within their interquartile ranges as determined by \tona. Finally, we selected only those abundance values that \bacc\ identified as good fits through its quality flag and with signal-to-noise ratios above 100 to ensure high-quality measurements. The Monte Carlo procedure produced, for each spectral feature, a distribution of up to one hundred abundance measurements. {We adopted 25 converged iterations as the minimum threshold for a star to be included in our final analysis. The final number of useful iterations varies from star to star, but about 85\% of the objects included in our analysis have more than 75 converged iterations.}

We adopted the median of the resulting distribution as the final abundance value, and the standard deviation as its error. {The distributions are approximately symmetric, although they are not strictly Gaussian, so the median is a robust estimator of the abundance, and the standard deviation provides a reasonable measure of its dispersion.}

To implement the latter procedure, we employed the GRID~UNAM\footnote{GRID UNAM is a high-performance computing infrastructure that connects various units of the National Autonomous University of Mexico (UNAM) and it provides both high-performance and high-throughput computing capabilities on heterogeneous clusters.} system, which enabled \bacc\ to be executed in a pseudo-parallel mode by distributing tasks across multiple threads within a networked server grid. When more than one line was available for a given element,
{the final abundance was computed as the weighted mean of the individual line abundances, $\epsilon(X) = \sum_i w_i\,\epsilon(X)_i / \sum_i w_i$, with weights $w_i = 1/\sigma_i^2$, where $\epsilon(X)_i$ and $\sigma_i$ are the median and standard deviation of the Monte Carlo distribution of the line $i$. The uncertainty of the final abundance is the error of the weighted mean, $\sigma_\epsilon = (\sum_i w_i)^{-1/2}$.} The final abundances and their associated uncertainties are reported in Table~\ref{tab:res}.

\subsection{Chemical Abundances}\label{sect:abun}
The use of accurate atmospheric parameters, such as those determined with \tona, together with the imposed low-\vsini\ criterion, increases the success rate of the chemical-abundance determination with \bacc.
Additional quality cuts (the \bacc\ quality flags, removal of problematic lines, and a minimum line signal-to-noise ratio) were essential to obtain robust abundances: without them, some stars show spurious abundances differing by up to 1 dex from solar values. Following these criteria, we derived abundances of C, N, O, and Fe for 226 stars ($\sim$62\% of the initial sample), while we were able to obtain S abundances for 173 objects ($\sim$48\% of the initial sample). Nevertheless, {we identified and marked 31 stars as outliers (see Appendix \ref{app:out} for details) that we removed for further analysis.} The final sample contains 195 stars: 20 in Orion A, 17 in Orion B, 118 in Orion OB, and 40 in $\lambda$~Ori, of which 161 have S abundances. Therefore, we determined the abundances of all five elements for 161 young stars that we discussed in the following sections.

\begin{table}
\caption{Basic information of the atomic/molecular lines used in this study. The last column shows if the line was considered for (Y)oung or (M)ain-sequence stars.} \label{tab:lines}
\begin{tabular}{ccccc}
\hline
{$\lambda$} & {$\chi$} & {$\log$ gf} & {Van der} & sample\\
{(\AA)} & {(eV)} & {} & {Waals} \\        
\hline
\multicolumn{5}{c}{CO features to compute C abundance} \\ \\
15578.048 & 0.251 & -7.588 & -- & Y \\
15784.709 & 6.999 & -5.592 & -- & M \\
16185.548 & 1.095 & -6.166 & -- & Y \\ 
16397.283 & 0.626 & -7.051 & -- & Y \\ 
16613.984 & 1.581 & -5.701 & -- & Y,M \\ 

\hline
\multicolumn{5}{c}{CN features to compute N abundance}\\ \\
15185.567 & 0.730 & -1.598 & -- & Y \\ 
15221.835 & 0.960 & -1.192 & -- & M\\
15228.739 & 0.836 & -1.968 & -- & Y \\
15308.892 & 1.012 & -1.179 & -- & Y,M \\
15418.090 & 0.921 & -1.783 & -- & Y \\
15580.910 & 0.896 & -1.680 & -- & Y \\
15636.594 & 1.131 & -1.271 & -- & Y,M \\
\hline
\multicolumn{5}{c}{OH features to compute O abundance}\\ \\
15568.782 &  0.299 & -5.337 & -- & M \\ 
15572.084 &  0.300 & -5.337 & -- & Y,M\\
16247.884 &  0.542 & -5.177 & -- & Y\\
16649.950 &  0.685 & -5.094 & -- & Y,M\\
16656.132 &  2.975 & -4.396 & -- & Y,M\\
16871.893 &  0.763 & -5.056 & -- & Y\\
 
\hline
\multicolumn{5}{c}{S {\sc i} atomic lines}\\ \\
15422.276 & 8.700 &  0.775 & -6.90 & Y \\
15478.482 & 8.045 & -0.040 & -6.84 & Y,M \\

\hline
\multicolumn{5}{c}{Fe {\sc i} atomic lines}\\ \\
15207.526 & 5.385 &  0.067  & -7.21 & Y\\
15244.974 & 5.587 & -0.134  & -7.01 & Y,M\\
15621.654 & 5.539 &  0.280  & -7.04 & M\\ 
15631.948 & 5.351 & -0.032  & -7.12 & M\\
15723.586 & 5.621 & -0.011  & -7.05 & Y,M\\
15904.324 & 6.365 & -0.154  & -7.19 & M \\
15920.642 & 6.258 & -0.011  & -7.01 & Y\\
15980.726 & 6.264 &  0.736  & -6.91 & M \\ 
16316.320 & 6.280 &  0.857  & -6.90 & Y\\
16645.874 & 5.956 & -0.226  & -6.90 & M \\
\hline
\end{tabular}
\end{table}

In Figure \ref{fig:location}, we show the spatial distribution of the stars of our sample color-coded by their respective abundance {of C, N, O, S, and Fe}, in the logarithmic scale $\epsilon$(X) = 12 + $\log$($n_x$/$n_H$). The five elements appear to be uniformly distributed throughout the analyzed region, with no obvious preferential spots, a result that reinforces the findings of \citetalias{lopezval25} regarding the chemical uniformity of the Orion star-forming region.

\begin{table}
\caption{Descriptive content of the Table of the stellar parameters and chemical abundances determined in this work. The full table is available in the electronic version of the paper.}
\label{tab:res}
\begin{tabular}{ll}
\hline
{Column name} & {Description} \\        
\hline
OBJ  &  2MASS identification number\\
RA   &  Right Ascension (J2000) \\
Dec  &  Declination (J2000) \\
Grp  &  Group of membership within Orion \\
t25  &  Effective temperature at percentile 25\\
t50  &  Effective temperature at percentile 50 \\
t75  &  Effective temperature at percentile  75\\
g25  &  Surface gravity at percentile 25 \\
g50  &  Surface gravity at percentile 50 \\
g75  &  Surface gravity at percentile 75\\
m25  &  Overall metallicity at percentile 25 \\
m50  &  Overall metallicity at percentile 50 \\
m75  &  Overall metallicity at percentile 75 \\
v25  &  Projected rotational velocity at percentile 25 \\
v50  &  Projected rotational velocity at percentile 50 \\
v75  &  Projected rotational velocity at percentile 75 \\

mean\_C &  Mean C abundance \\
erC &  Error in C abundance \\
lC  &  Lines used in C abundance determination \\

mean\_N &  Mean N abundance \\
erN &  Error in N abundance \\
lN   &  Lines used in N abundance determination \\

mean\_O&  Mean O abundance \\
erO&  Error in O abundance \\
lO  &  Lines used in O abundance determination \\

mean\_S &  Mean S abundance \\
erS&  Error in S abundance \\
lS  &  Lines used in S abundance determination \\

mean\_Fe &  Mean Fe abundance \\
erFe &  Lower error in Fe abundance \\
lFe   &  Lines used in Fe abundance determination \\
flag\_out & Outlier flag \\
\hline
\end{tabular}
\end{table}

\subsubsection{Carbon}
{C is synthesized in stars through the triple-alpha reaction during He burning. Part of this C is then converted into O  so the
competition between these two reactions determines the C/O ratio at the end
of core He burning \citep{arconesthielmann2023}. About half of the C in the
Universe is produced by massive stars (M\,$>$\,10\,M$_\odot$), while the rest
is mainly produced by low-mass AGB stars
(1--4\,M$_\odot$). Because these AGB stars do not produce new Fe, 
they enhance the [C/Fe] ratio efficiently, and this ratio is expected to decrease
once Type~Ia supernovae begin to contribute Fe \citep{kobayashikarakaslugaro2020}.}

{Although {in} \citetalias{lopezval25} the C abundance was obtained for 280 stars, we decided to recompute {it here}, as there is abundance coupling between the molecular features used, meaning that variations in the abundance of one element modify the molecular number densities and therefore affect the abundances derived from the other molecular species.}
This set of C abundances {is} on average 0.11$\pm$0.15 dex lower than those of \citetalias{lopezval25}.

To account for possible differences in the chemical behavior of stars with different spectral types, we split our sample into two temperature bins: M (\teff~$<$~4000~K) and K (4000~$\leq$~\teff~$\leq$~5000~K) stars. According to these temperature cuts, our {final} sample comprises {59} M{-type} stars, {134} K{-type} stars, and {2} stars with \teff~$>$~5000~K (mostly G-type sources). {We found that median C abundances are comparable between the {temperature} cuts (see Table~\ref{tab:mk_comparison}). This result shows that the selected CO features used to determine C abundance are suitable to {characterize} a wide range of temperatures, as there is {no} significant difference as a function of temperature.}

We adopted the published solar standard scale of \cite{asplund21} in order to compare our results with the Sun {and to obtain [X/H]\footnote{[X/H] = $\epsilon(\rm X)_{\rm star}$ - $\epsilon({\rm X})_\odot$} abundance ratios}. It is important to note that all the conclusions on such comparisons are sensitive to the adopted solar reference scale, although they can be straightforwardly extended to other standard abundance scales if required. {Throughout this work, the uncertainty of the median [X/H] combines in quadrature the sample MAD and the solar abundance uncertainty, and we express the significance of any departure from the solar value in units of this total uncertainty (hereafter ``$\sigma$'')}

We obtained a median C abundance for the whole sample of young stars (see Table~\ref{tab:mk_comparison}) {that is consistent within the errors with the solar value. On the other hand, the median abundance of K-type stars, given in Table~\ref{tab:mk_comparison}, is consistent with the solar value, while for the M-type stars it is only marginally consistent, differing from solar by about 1.8$\sigma$.}


\begin{table*}
\centering
\caption{Median abundances and [X/H] ratios for the whole sample of Orion stars and main-sequence stars of the solar neighborhood analyzed here. We also report the median abundance values for the M- and K-type star subsamples. For reference, we adopted the solar value of \citet{asplund21}, used to compute [X/H], which is reported in the last column.}
\label{tab:mk_comparison}
\begin{tabular}{l c c c c c c c}
\hline\hline
& \multicolumn{2}{c}{\teff$\le$4000 K} & \multicolumn{2}{c}{4000 $<$ \teff\ $\le$5000 K} & \multicolumn{2}{c}{whole sample} & \\
Quantity & Orion stars& Main-sequence stars & Orion stars& Main-sequence stars& Orion stars & Main-sequence stars& Solar value \\
\hline
$\epsilon$(C) & 8.28$\pm$0.09 & 8.42$\pm$0.10 & 8.42$\pm$0.07 & 8.51$\pm$0.11 & 8.37$\pm$0.10 & 8.48$\pm$0.13 & \multirow{2}{*}{8.46$\pm$0.04} \\
$[\mathrm{C/H}]$         & $-$0.18$\pm$0.10 & $-$0.04$\pm$0.08 & $-$0.04$\pm$0.11 & $+$0.05$\pm$0.12 & $-$0.09$\pm$0.11 & 0.02$\pm$0.13\\ \\

$\epsilon$(N) & 7.36$\pm$0.22 & 7.48$\pm$0.21 & 7.34$\pm$0.14 & 7.51$\pm$0.21 & 7.35$\pm$0.16 & 7.56$\pm$0.23 & \multirow{2}{*}{7.83$\pm$0.07} \\
$[\mathrm{N/H}]$         & $-$0.47$\pm$0.23 & $-$0.35$\pm$0.22 & $-$0.49$\pm$0.16 & $-$0.32$\pm$0.22 & $-$0.48$\pm$0.17 & $-$0.27$\pm$0.24 & \\ \\

$\epsilon$(O) & 8.59$\pm$0.10 & 8.67$\pm$0.08 & 8.65$\pm$0.07 & 8.74$\pm$0.11 & 8.63$\pm$0.08 & 8.73$\pm$0.12 & \multirow{2}{*}{8.69$\pm$0.04} \\
$[\mathrm{O/H}]$         & $-$0.10$\pm$0.11 & $-$0.02$\pm$0.09 & $-$0.04$\pm$0.08 & $+$0.05$\pm$0.12 & $-$0.06$\pm$0.09 & $+$0.04$\pm$0.12 \\ \\

$\epsilon$(Fe) & 7.45$\pm$0.05 & 7.35$\pm$0.09 & 7.45$\pm$0.04 & 7.43$\pm$0.09 & 7.45$\pm$0.05 & 7.43$\pm$0.10 & \multirow{2}{*}{7.46$\pm$0.04} \\
 $[\mathrm{Fe/H}]$         & $-$0.01$\pm$0.06 & $-$0.11$\pm$0.10 & $-$0.01$\pm$0.06 & $-$0.03$\pm$0.10 & $-$0.01$\pm$0.06 & $-$0.03$\pm$0.11 \\ \hline
No. stars & 59 & 46 & 134 & 286 & 195 & 430 & \\ \\  

$\epsilon$(S) & -- & --& 7.19$\pm$0.19 & 7.14$\pm$0.16 & 7.19$\pm$0.18 & 7.12$\pm$0.14 & \multirow{2}{*}{7.12$\pm$0.03} \\
$[\mathrm{S/H}]$ & -- & --& $+$0.07$\pm$0.19 & $+$0.02$\pm$0.16 & $+$0.07$\pm$0.18 & 0.00$\pm$0.15\\ \hline
No. stars & 0 & 0 & 126 & 234 & 128 & 330 & \\ \\

\hline
\hline

\end{tabular}

\end{table*}

\subsubsection{Nitrogen}     
{N can be synthesized through both primary and secondary production
channels. Primary nitrogen is generated independently of the initial stellar
metallicity, whereas secondary N is produced through CNO processing of
pre-existing C and O and therefore increases with metallicity
\citep{romanomatteuccizhang2019, grisonimatteucciromano2021}. The latter
contribution is mainly associated with intermediate-mass stars, whose
evolutionary timescales ($\sim$40 Myr) are comparable to the minimum delay
times required for the onset of Type Ia supernovae, the main producers of iron
\citep{pakmormattheematsuno2024}.}

The N abundance was derived {from three or
more spectral features for the $\sim$88\% (170 of 195 stars) of our final sample. 
The 25 stars with only two lines are predominantly objects with \teff\ $<$ 4000~K 
(22 of 25)}. Cooler stars are intrinsically more difficult to characterize due to
their stronger molecular blends and lower signal-to-noise ratio spectra, so
these measurements should be interpreted with caution.

{Our results show that the median N abundance of the whole sample is subsolar at the level of $\sim$2.8$\sigma$. Regarding the M-type and K-type stars, we found that both subsamples have similar median N abundance, also sub-solar at the level of $\sim$2$\sigma$ and $\sim$3$\sigma$, but the associated total uncertainty is 40\% larger in the M-type stars, which might be related to the spectral complexity of these objects (see Table \ref{tab:mk_comparison}). }

\subsubsection{Oxygen}
{O is primarily produced through hydrostatic nucleosynthesis in the interiors of massive stars with masses greater than 8 M$_{\odot}$, following He burning via the triple-alpha process that forms C and subsequently heavier elements. The interstellar medium becomes enriched in oxygen through stellar winds during the late evolutionary phases and through the explosions of dying stars as core-collapse supernovae on short timescales (e.g. \citealp{chiappiniromanomatteucci2003}).}

We found the median O abundance consistent, within the errors, with the solar value.
About {76\% (149 out of 195)} of our O abundances were derived from averaging four or more OH lines. The OH lines have long been considered among the most reliable oxygen abundance indicators \citep[e.g.][]{grevesse84} and have been extensively used, particularly in cases where atomic O lines are too weak, blended, or inaccessible, such as in cool and solar-like stars ($\sim$3500\,K $\lesssim T_{\rm eff} \lesssim$ 6000\,K) and metal-poor stars \citep[e.g.][]{melendez01,garciaperez06,holtzman15,jonsson20,meszaros25}. If we compare our median O abundances between M{-type} and K{-type} stars, both values (Table~\ref{tab:mk_comparison}) are very similar and consistent with each other.

For the O abundances, we did not find a large dispersion in our measurements among the M stars, as was the case for N. This could be attributed to the derivation of the O abundance from OH lines, which are more prominent in cooler stars, and {might be less blended compared to those CN features}.

\subsubsection{Iron}
{Fe is mainly produced by Type~Ia supernovae, explosions of C+O white dwarfs in binary systems, with a smaller contribution
from core-collapse supernovae of massive stars \citep{kobayashikarakaslugaro2020}. In these explosions, the Fe is synthesized from the decay of radioactive $^{56}$Ni \citep{arconesthielmann2023}. Type~Ia supernovae produce on average about 
0.6\,M$_\odot$ of $^{56}$Ni, compared with about 0.1\,M$_\odot$ from core-collapse supernovae. Although core-collapse events are more frequent, by a factor of 4--5, SNe~Ia are the dominant producers of Fe, contributing about 50--55\% of the Fe in
the solar system, with the remainder coming from core-collapse supernovae \citep{arconesthielmann2023}.}

Compared to those Fe values determined in \citetalias{lopezval25}, we found a median difference of 0.00$\pm$0.06 dex between both sets of abundances. We found a median Fe abundance in good agreement with the solar value. In terms of M- and K-type stars, we did not find any difference between them (Table~\ref{tab:mk_comparison}); both values are also consistent with the solar value.

\subsubsection{Sulphur}
{S shares a common nucleosynthetic origin with O; both are $\alpha$
elements produced mainly in massive stars and ejected by core-collapse
supernovae, and therefore their abundance ratio is expected to remain
approximately constant. However, some chemical evolution models, such as
those presented by \citet{kobayashikarakaslugaro2020} and
\citet{arconesthielmann2023}, predict an additional contribution to S from
Type~Ia supernovae. This issue remains under debate, mainly because of
observational limitations.}

S has {a smaller} number of measurements in our study, with {161}. This is because all the atomic S lines analyzed show strong trends between $\epsilon$(S) and \teff\ at temperatures below $\sim$4000 K, which limits our ability to determine S abundances in cooler stars. This behavior reflects the intrinsic range of applicability of these atomic lines. The S abundances derived for stars with \teff\ $\lesssim$ 4000 K are less reliable than those of hotter stars, and should be used with caution. For this reason, we only considered the 128 stars with \teff\ $\gtrsim$ 4000 K in our analysis.
We found a median S abundance (see Table~\ref{tab:mk_comparison}) for the whole sample consistent, within the errors, with the solar value of \cite{asplund21}. 


\subsection{Orion chemical homogeneity}\label{sect:grp_var}

{In this section, we search for chemical inhomogeneities in the Orion complex, both between its groups and within each of them. If present, such variations could point to localized enrichment events in the past, which should be consistent with the average age of each group.}

\subsubsection{Age of the Orion groups}\label{sect:age}

{
To associate the chemical abundances derived in this work with the ages of the stellar populations across the Orion complex, we cross-matched our stellar sample with the catalogue of young stellar structures identified by \citet{sanchez-sanjuan:2024}. The catalogue was constructed using \textit{Gaia} DR3 astrometric and photometric data to select pre-main-sequence star candidates, which were subsequently grouped into spatially and kinematically coherent structures using a clustering algorithm applied to the five-dimensional astrometric space defined by sky position, parallax, and proper motions. The resulting groups were classified into two categories according to their stellar membership: Big Structures and Small Structures. Our cross-match yielded 129 sources associated with 12 Big Structures and eight additional sources belonging to four Small Structures.}

{
For the Big Structures, we adopted the ages derived by \citet{sanchez-sanjuan:2026}, who fitted PARSEC \citep{parsecmarigo} evolutionary isochrones to distance-corrected \textit{Gaia} DR3 CMDs. The fitting procedure followed a maximum-likelihood approach, treating age and visual extinction as free parameters while assuming a fixed solar metallicity. Age uncertainties were estimated from 100 Monte Carlo realizations, with the 16th and 84th percentiles of the resulting distributions defining the lower and upper uncertainties, respectively. Each cross-matched star was assigned the age of its parent structure rather than an individually determined stellar age. For OBP-near (C10), we adopted an age of $\sim$8.7 Myr, calculated as the arithmetic mean of the ages derived for its two components, OBP-near A and OBP-near B. For the Small Structures OBP-South (Cl) and OBP-Centre (Ck), we adopted the ages reported by \citet{sanchez-sanjuan:2024}, estimated using an approximate CMD-based method. No ages were available for the remaining Small Structures, L1616 A (Cf) and OBP-West (Ce).}

{
To establish the correspondence with the four Orion regions considered in this work, we associated ONC and Ori-South with Orion A, Ori-East and $\sigma$ Ori with Orion B, and $\lambda$ Ori with the $\lambda$ Orionis region. The remaining structures were assigned to Orion OB. For each region, we calculated a representative age as the weighted mean, $\overline{\tau}$, of the individual substructure ages, using the number of cross-matched sources in each substructure as weights. Structures without available age estimates were excluded from this calculation. The uncertainties of the weighted mean ages were obtained by propagating the asymmetric uncertainties of the individual substructure ages, assuming independent age estimates. Table~\ref{tab:orion_substructure_ages} summarizes the correspondence between the stellar structures and the four Orion regions, together with their cross-matched source counts, individual ages, and representative regional ages.}

{According to the ages of \citet{sanchez-sanjuan:2026} (reported in
Table~\ref{tab:orion_substructure_ages}), Orion A and B have very similar
ages of 3--4~Myr, {which lie at the upper end of} the typical ages of $\sim$2--4~Myr
reported previously \citep[e.g. ][]{flaherty08,dario10,kounkel18,roman-zuniga23},
and are the youngest groups in our study. We estimate an average age of
$\sim$7~Myr for $\lambda$~Ori, which is {also at the upper end of} previous
determinations \citep[$\sim$2--7~Myr;][]{bayo11,zari19,cao22,roman-zuniga23}.
Finally, the oldest group in our study is Orion OB. As can be seen in
Table~\ref{tab:orion_substructure_ages}, it contains a large number of the
substructures of \citet{sanchez-sanjuan:2026}, but most of them agree on an
older age of about 11~Myr, which is slightly higher than, but still in
agreement with, previous determinations
\citep[$\sim$4--10~Myr;][]{briceno07,bally08,downes14,roman-zuniga23}.
{The ages of \citet{sanchez-sanjuan:2026} thus tend to be on the older side of previous estimates. This is not unexpected, since the absolute ages derived from pre-main-sequence isochrones depend on the adopted evolutionary models; for instance, \citet{dario10} find that the age distribution of the ONC peaks between $\sim$2 and $\sim$3~Myr depending on the models, and \citet{rottens26} report that absolute ages in Orion depend on the isochrone model at the level of 1--3~Myr. Since the relative ages of the groups and the total age spread are much less sensitive to a systematic offset in the absolute scale, our conclusions are not affected by this.}}

\begin{table}
\centering
\renewcommand{\arraystretch}{1.26}
\caption{Ages of the four Orion groups considered in this work. The groups are matched to the stellar structures identified by \citet{sanchez-sanjuan:2024}, whose identifiers (ID) and substructure names are adopted here. Ages and uncertainties for the Big Structures are taken from \citet{sanchez-sanjuan:2026}, and those of the Small Structures Cl and Ck from \citet{sanchez-sanjuan:2024}. The representative age of each group is the mean of its substructure ages, weighted by $N_{\rm match}$.}
\label{tab:orion_substructure_ages}
\begin{tabular}{lllccc}
\hline\hline
Group &
ID &
Substructure &
$N_{\rm match}$ &
Age (Myr) &
\shortstack{$\overline{\tau}$ (Myr)} \\
\hline

\multirow{2}{*}{Orion A}
 & C11 & ONC
 & 11 & $4.1^{+0.6}_{-2.1}$
 & \multirow{2}{*}{$4.0^{+0.4}_{-1.5}$} \\
 & C12 & Ori-South
 & 4 & $3.6^{+0.6}_{-0.5}$ & \\
\hline

\multirow{2}{*}{Orion B}
 & C05 & Ori-East
 & 2 & $3.1^{+1.1}_{-0.3}$
 & \multirow{2}{*}{$3.2^{+0.5}_{-0.7}$} \\
 & C07 & $\sigma$ Ori
 & 4 & $3.3^{+0.5}_{-1.1}$ & \\
\hline

$\lambda$ Ori
 & C01 & $\lambda$ Ori
 & 39 & $6.8^{+0.8}_{-1.5}$
 & $6.8^{+0.8}_{-1.5}$ \\
\hline

\multirow{11}{*}{Orion OB}
 & C02 & Ori-North
 & 3 & $18.1^{+1.7}_{-1.0}$
 & \multirow{11}{*}{$11.4^{+0.4}_{-0.4}$} \\
 & C03 & Brice\~no-1A
 & 4 & $11.0^{+0.8}_{-0.3}$ & \\
 & C04 & Brice\~no-1B
 & 17 & $12.4^{+1.2}_{-1.0}$ & \\
 & C06 & OBP-far
 & 4 & $15.7^{+1.1}_{-0.3}$ & \\
 & C08 & OBP-b
 & 3 & $18.9^{+0.6}_{-0.5}$ & \\
 & C09 & OBP-d
 & 15 & $11.0^{+1.0}_{-0.9}$ & \\
 & C10 & OBP-near
 & 23 & $8.7^{+0.6}_{-0.7}$\,$^{\dagger}$ & \\
 & Cf & L1616 A
 & 4 & -- & \\
 & Ce & OBP-West
 & 2 & -- & \\
 & Cl & OBP-South
 & 1 & $10.3^{+2.3}_{-1.3}$ & \\
 & Ck & OBP-Centre
 & 1 & $3.5^{+2.5}_{-1.9}$ & \\
\hline\hline
\end{tabular}
\end{table}

\subsubsection{Abundance dependence on atmospheric parameters}
{Although we followed a strict procedure to select the most suitable atomic/molecular lines to analyse our Orion sample, namely those lines with a weak correlation with \teff\ and \logg\ ($r_{\rm s}\leq0.4$), and performed a Monte Carlo to propagate uncertainties, the chemical abundances may still depend on the atmospheric parameters. This is because the final abundances come from a weighted average of different lines, and different stars may not share the same lines in that average, and also because the Orion groups might not sample the atmospheric parameters identically.}

{To quantify how our chemical abundances depend on the atmospheric parameters, we performed a multiple linear regression for each element as in which the abundance is described as a constant plus a linear combination of \teff, \logg, \meta, and \vsini\ as follows: }

\begin{equation}
\epsilon({\rm X}) = \alpha_0\; + \; \alpha_1\,$\teff$\; +\; \alpha_2\,$\logg$\; + \;\alpha_3$\,\meta$\; +\; \alpha_4$\,\vsini$
\label{eq:regression}
\end{equation}

{We preferred this approach over fitting each parameter separately because in \bacc\ the complete set of atmospheric parameters is treated as the true values, and they might not be fully independent of each other \citep[e.g.; ][]{hinkel16, jofre19}. In the multiple regression, each coefficient tells us how much the derived abundance changes when its parameter varies while the other three are kept fixed. Table~\ref{tab:param_dependence} lists the coefficients and their uncertainties per unit of each parameter (100~K, 0.1~dex, or 1~km~s$^{-1}$), whereas our stars span a much wider range. To assess their effect on the abundances, it is more informative to multiply each coefficient by the range covered by the central 68\% of our sample in the corresponding parameter (about 590~K in \teff, 0.38~dex in \logg, 0.09~dex in \meta, and 9.6~km~s$^{-1}$ in \vsini). With this, the dependence on the parameters changes the abundances by up to $\sim$0.16~dex, and in most cases between 0.03 and 0.15~dex, which is comparable to the dispersion of each abundance within the sample (MAD between 0.05 and 0.18~dex). Therefore, the dependence on the atmospheric parameters is not negligible when comparing the abundances of groups that do not share the same parameters, and it must be considered before attributing a difference between groups to a real chemical variation.}

\begin{table}
\centering
\caption{Dependence of the derived abundances on the atmospheric parameters, obtained from a multiple linear regression in which the four parameters are fitted simultaneously. $\beta_0$ is the abundance (in dex) predicted for a reference star with the median values of the sample (\teff = 4150 K, \logg = 4.08, \meta = -0.08, and \vsini = 19.2 km s$^{-1}$); the same reference is used for all the elements, including S, whose subsample (\teff $> 4000$\,K) has slightly different median values. The remaining coefficients and their uncertainties are given in units of $10^{-3}$\,dex per unit of the parameter indicated in the header.}
\label{tab:param_dependence}
\setlength{\tabcolsep}{4pt}
\begin{tabular}{lccccc}
\hline\hline
El. & $\alpha_0$ & $\alpha_1$ & $\alpha_2$ & $\alpha_3$ & $\alpha_4$ \\
 & (dex) & ($10^{-3}$\,dex & ($10^{-3}$\,dex & ($10^{-3}$\,dex & ($10^{-3}$\,dex \\
 &       & per 100\,K)     & per 0.1\,dex)   & per 0.1\,dex)   & per km s$^{-1}$) \\
\hline
C  & $8.38\pm0.01$ & $11\pm3$   & $19\pm8$   & $59\pm21$  & $11\pm3$ \\
N  & $7.32\pm0.02$ & $27\pm6$   & $1\pm14$   & $110\pm37$ & $-10\pm5$ \\
O  & $8.64\pm0.01$ & $17\pm2$   & $19\pm6$   & $35\pm15$  & $8\pm2$ \\
S  & $7.23\pm0.03$ & $-15\pm12$ & $40\pm22$  & $-56\pm57$ & $-5\pm7$ \\
Fe & $7.45\pm0.01$ & $0\pm2$    & $-6\pm4$   & $78\pm9$   & $0\pm1$ \\
\hline
\end{tabular}
\end{table}

\subsubsection{Chemical homogeneity across the complete complex}\label{sect:homo}
{In Figure \ref{fig:location}, we show the spatial distribution of our final sample color-coded by the five elemental chemical abundances analysed here. To assess whether the abundances of the four Orion groups are statistically distinguishable, we restricted the analysis to a small set of simple, non-parametric tests, which make no assumption about the shape of the abundance distributions and are valid for groups of different sizes, as is the case here: our final sample of 195 stars is distributed as 20, 17, 118, and 40 stars in Orion~A, Orion~B, OB, and $\lambda$~Ori, respectively. In all of these tests, the null hypothesis is that the stars being compared are drawn from the same underlying distribution, and the $p$-value is the probability of finding a difference at least as large as the observed one if that hypothesis were true. A small $p$-value (we adopt $p< 5\%$) therefore indicates that the groups are unlikely to be identical. All tests were performed with the \textsc{SciPy} package \citep{virtanen20}.}

\begin{figure*}
    \centering
    \includegraphics[width=\linewidth]{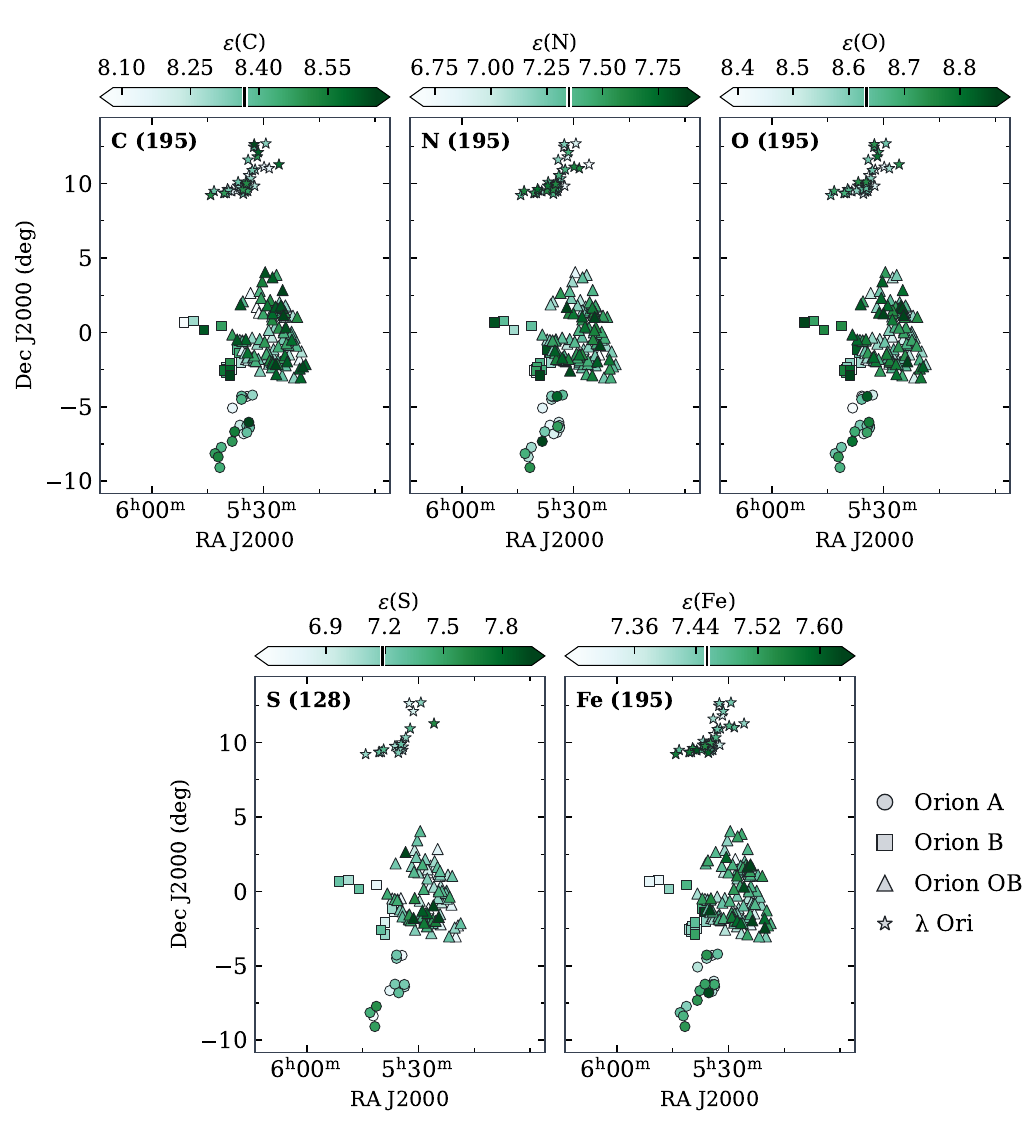}
    \caption{Spatial distribution of $\epsilon$(X) abundances across the observed Orion's subgroups. Each point corresponds to an individual star with reliable abundance measurements (i.e., excluding outliers), while different symbols denote stars belonging to different regions of Orion. The color scales illustrate the abundance range spanned by the stellar sample. In each color bar, the median abundance value of the whole complex is indicated.}
    \label{fig:location}
\end{figure*}

{We started applying a Kruskal-Wallis test \citep{kruskal52}. In this test, all the stars of the four groups are ranked from the lowest to the highest abundance value. The test then checks whether the ranks of one or more groups tend to be systematically higher or lower than those of the others. It is a single test per element. For C, Fe, and S, we find no significant differences between the groups ($p= 19\%$, 35\%, and 45\%, respectively). In contrast, the Kruskal-Wallis test indicates differences for N ($p\sim3\%$) and O ($p=0.06\%$).}

{According to the later statistical test, our groups have some differences in N and O. To identify which group (or groups) is different, we applied a Mann-Whitney U test \citep[M-W; ][]{mann47}. The M-W test compares two groups and evaluates how likely it would be, purely by chance, to find a difference in the abundances at least as large as the observed one if the two groups were identical. Because we have 4 different groups, there are six possible pairs of groups to analyse in N and O. Since each element involves six pairwise comparisons, some of them could yield a small $p$-value purely by chance. We applied a Holm correction \citep{holm79} to adjust the $p$-values ($p_{\rm cor}$) to account for the number of comparisons performed.}

{For N, the difference is driven by Orion~A, whose median is lower than those of the three remaining groups (see Table~\ref{tab:orion_groups}). The A--OB pair shows the most significant difference, with a Holm-corrected Mann-Whitney $p_{\rm cor}=1.2\%$. The difference between the medians of the A and OB groups is 0.29~dex, with Orion~A being the one with the lowest N. We compare through a Kolmogorov-Smirnov (K-S) test \citep{kolmogorov33, smirnov48} the atmospheric parameter distributions (\teff, \logg, \meta, and \vsini) of both groups. After a Holm correction for 4 tests, the K-S test showed no significant differences (at the 5\% level) between the two groups in any of the parameters. The $p_{\rm cor}$-values are $\sim$16\% for \teff, \meta, and \vsini, and 5.6\% for \logg, the closest to the significance threshold. Orion~A has a median \logg\ 0.13~dex lower than that of OB. If we consider \logg\ as a proxy for stellar age, higher values might translate to older ages; although not significant, this result is in line with the difference in age of the groups, with Orion OB (see Table~\ref{tab:orion_substructure_ages}) being older than Orion~A. However, following Table~\ref{tab:param_dependence}, the differences in the median atmospheric parameters between the two groups ($\sim$96~K in \teff, $\sim$0.13~dex in \logg, $\sim$0.01~dex in \meta, and $\sim$0.1~km~s$^{-1}$ in \vsini) might produce a difference in the N abundances of only $\sim$0.04~dex. Thus, the dependence of the derived abundances on the atmospheric parameters cannot account for the difference in N between A and OB.}

{As we can see in Table~\ref{tab:orion_groups}, the groups B, OB, and $\lambda$~Ori have subsolar [N/H] values of $\sim$$-0.5$ ($-0.51$ to $-0.46$), and Orion~A has $-0.75$, whereas the [X/H] ratios for C, O, S, and Fe are solar or close to solar ($|{\rm [X/H]}| \lesssim 0.12$~dex). For nearby young stars, we expected an N abundance close to the solar value, as [N/Fe] remains close to solar at solar metallicity in nearby disk dwarfs \citep[e.g.][]{reddy03,bensby26}. Thus, our N abundances might present a systematic bias toward lower values, but Orion~A appears to have an additional deficit against the other three groups that makes it statistically distinct. The cause of this deficit is not clear; if real, the stars of Orion~A might have formed from an N-depleted gas distinct from the whole complex, as there is not enough time to let the CNO cycle change the surface N abundance in these young stars. The fact that there is a negative offset from the other elements might point to a systematic origin and require further analysis. The CN features are more sensitive to \teff\ and \logg, and to the formation of CO and OH \citep[e.g.][]{russell34,tsuji73}. Additionally, the Orion~A stars are younger and might have more stellar activity, spots, and higher magnetic fields, which complicate measuring the molecular lines.}

{Since the abundances are derived iteratively (Fe, O, C, N, and Fe again until convergence, after which S is computed), a bias in N could in principle propagate to the other elements, for example through its effect on the molecular equilibrium of the atmosphere. We do not find evidence of this in our results. The median abundances of Orion~A differ from those of OB by only $-0.015$, $-0.038$, $0.012$, and $-0.014$~dex for C, O, S, and Fe, respectively, compared to $-0.29$~dex for N, and none of these differences is significant (Holm-corrected Mann-Whitney $p_{\rm cor} \geq 45\%$ for all the pairs involving Orion~A). Fe is of particular interest because it is recomputed after N in the last step of the iterative process, and it shows no offset in Orion~A. In addition, the [C/O] ratio is consistent across all the groups (median values between $-0.05$ and $0.02$, with a MAD of 0.05--0.07 in each group, and $-0.03 \pm 0.06$ for the whole sample), and [N/Fe] is the only abundance ratio relative to Fe that remains significant after the Holm correction.}

{In summary, we find no evidence that the N deficit of Orion~A affects the abundances of C, O, S, and Fe derived for the same stars. The difference in N between Orion~A and OB is statistically significant. The dependence of the derived abundances on the atmospheric parameters, evaluated for the full sample, does not account for why only Orion A is affected. Whether this is a real effect or an artifact of the analysis cannot be determined with the tests performed here, and further work may be required.}

{Regarding O, the pair distinct is OB--$\lambda$-Ori, as their Holm-corrected M-W test gives a $p_{\rm cor}$-value of 0.74\%. The difference in the median O abundance between the two groups (Table~\ref{tab:orion_groups}) is $\sim$0.05~dex, and, as in the case of N, we compare the atmospheric parameter distributions of both groups through a K-S test. We found that the atmospheric parameter distributions of OB and $\lambda$-Ori are statistically indistinguishable for \logg, \meta, and \vsini. The Holm-corrected K-S $p$-value for the \teff\ distribution comparison is of $\sim$1\%, with the stars of $\lambda$-Ori being, on median, 200~K cooler than those of OB. }

{Following the results of the multiple linear regression (equation~\ref{eq:regression} and Table~\ref{tab:param_dependence}), a difference of $\sim$200~K in \teff\ might produce a difference in the O abundances of about 0.034~dex, which is a large fraction ($\sim$68\%) of the difference found in the comparison of the median O abundances. The differences between the two groups in the other three parameters are small ($\sim$0.03~dex in \logg, $\sim$0.01~dex in \meta, and $\sim$1.5~km~s$^{-1}$ in \vsini) and, with their respective coefficients, they add up to about 0.02~dex. Thus, the differences in the atmospheric parameters are of the order of the difference in O between the two groups, and we do not consider it as evidence of a significant chemical difference between the two Orion groups.}

{In conclusion, Orion is chemically homogeneous in C, O, S, and Fe {within the uncertainties of our analysis and for the stars studied here, although subtle differences may exist below our current detection limits, or in regions and stars not explored in this work.} This lack of significant abundance differences is consistent with the small age spread of the groups analysed, of about 8~Myr from the youngest to the oldest group (see Table~\ref{tab:orion_substructure_ages}), a period that is very short compared with the timescales (tens of Myr to Gyr) over which the interstellar medium is enriched by the bulk of stellar populations \citep[e.g.][]{kobayashikarakaslugaro2020}. For N, the groups B, OB, and $\lambda$~Ori are homogeneous, whereas Orion~A shows a lower N abundance. This difference cannot be explained by the dependence of the derived abundances on the atmospheric parameters, nor does it follow an age sequence: Orion~B, which is slightly younger than Orion~A, is consistent in N with the older groups $\lambda$~Ori and OB. A larger sample of stars in Orion~A is required to determine the origin of the N depletion.}

\begin{table*}
\centering
\caption{Median abundances and [X/H] abundance ratios for the four Orion groups analyzed in this work. For reference, we report the solar value of \citet{asplund21} in the last column, which we used to obtain the [X/H] abundance ratios.}
\label{tab:orion_groups}
\begin{tabular}{lccccc}
\hline\hline
& \multicolumn{4}{c}{Orion group} & \\
Quantity & A & B & OB & $\lambda$ Ori & Solar value\\
\hline
$\epsilon$(C) & 8.37$\pm$0.09 & 8.41$\pm$0.15 & 8.39$\pm$0.09 & 8.35$\pm$0.10&\multirow{2}{*}{8.46$\pm$0.04} \\
$[\mathrm{C/H}]$& -0.09$\pm$0.10 & -0.05$\pm$0.15 & -0.07$\pm$0.10 & -0.11$\pm$0.11 &  \\ \\

 $\epsilon$(N) & 7.08$\pm$0.17 & 7.32$\pm$0.15 & 7.37$\pm$0.15 & 7.36$\pm$0.16 & \multirow{2}{*}{7.83$\pm$0.07} \\
$[\mathrm{N/H}]$ & $-$0.75$\pm$0.18 & $-$0.51$\pm$0.17 & $-$0.46$\pm$0.17 & $-$0.47$\pm$0.17 &   \\ \\

$\epsilon$(O) & 8.61$\pm$0.10 & 8.70$\pm$0.11 & 8.65$\pm$0.08 & 8.61$\pm$0.08 & \multirow{2}{*}{8.69$\pm$0.04} \\
$[\mathrm{O/H}]$ & $-$0.08$\pm$0.11 & 0.01$\pm$0.12 & $-$0.04$\pm$0.09 & $-$0.08$\pm$0.09 &  \\ \\

$\epsilon$(Fe)     & 7.45$\pm$0.06 & 7.43$\pm$0.06 & 7.46$\pm$0.04 & 7.45$\pm$0.04 & \multirow{2}{*}{7.46$\pm$0.04} \\
 $[\mathrm{Fe/H}]$ & $-$0.01$\pm$0.07 & $-$0.03$\pm$0.07 & 0.00$\pm$0.06 & $-$0.01$\pm$0.06 &  \\
 \hline
No. stars   & 20 & 17 & 118 & 40 &   \\ \\
$\epsilon$(S)  & 7.22$\pm$0.22 & 7.10$\pm$0.19 & 7.21$\pm$0.20 & 7.12$\pm$0.14 & \multirow{2}{*}{7.12$\pm$0.03} \\
  $[\mathrm{S/H}]$ & 0.10$\pm$0.22 & $-$0.02$\pm$0.19 & 0.09$\pm$0.21 & 0.00$\pm$0.14 & \\
  \hline
No. stars  & 12& 9& 88& 19&  \\ \\

\hline
\hline
\end{tabular}
\end{table*}

\subsubsection{Chemical homogeneity within Orion groups.}
{In the previous section, we tested the chemical homogeneity of our sample as a whole Orion complex, while in the following we will focus on the four different groups and look for more local abundance variations. We computed the $r_s$ and its probability between the chemical abundance and Right Ascension (RA) and Declination (Dec). Later, as a complement, we implemented a Moran's $I$ test \citep{moran50} of nearby neighbors to determine whether stars that are close on the sky tend to have more similar abundances than expected by chance. For each star, we considered its five nearest neighbors in the plane of the sky and compared its abundance with theirs: a positive $I$ indicates that neighboring stars are more alike than the rest, while a value close to the expected one ($I= -1/n-1$) for a random distribution indicates no spatial structure. We consider a result as significant when $p_{\rm cor}$-value is lower than 5\%.} 

\textit{$\lambda$-Ori: } {We found weak correlations (according to \citealt{evans96}) between the N abundance and RA ($r_s=+0.38$) and between the Fe abundance and both RA ($r_s=+0.36$) and Dec ($r_s=-0.36$), with uncorrected $p$-values of 1.7\% for N, and 2.1\% and 2.4\% for Fe, respectively. However, after applying the Holm correction for the ten tests performed in this group (five tests per coordinate), none of them remain significant at the 5\% level ($p_{\rm cor} \geq 17\%$ in the three cases). The Moran test shows no evidence of spatial structure, as all the $p$-values are above 5\%, with O being the closest to this threshold ($p = 7\%$). }

{Although some authors have claimed the existence of a supernova $\sim$1--2~Myr ago \citep[e.g. ][]{maddalena87,cunha96,dolan02,kounkel18}, we did not find any evidence for chemical inhomogeneities in $\lambda$-Ori that can be associated with this past event. Given the achieved precision in our chemical abundances and the limited number of stars (40), we conclude that $\lambda$-Ori is consistent with being chemically homogeneous in the analysed elements, in agreement with previous studies \citep[e.g. ][]{biazo11,kos21}. In Figure~\ref{fig:lori} we show the spatial distribution of the stars of $\lambda$-Ori, color-coded by the difference between their abundance and the median abundance of the group, for each element.}

\begin{figure*}
\centering
\includegraphics[width=\textwidth]{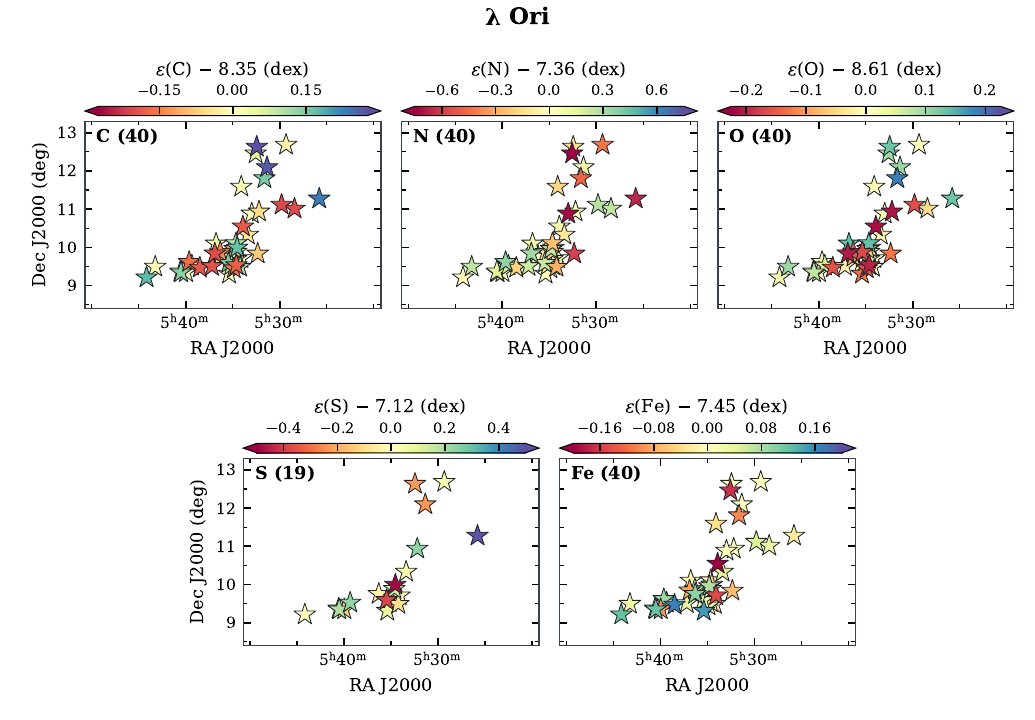}
\caption{Spatial distribution of the stars in the $\lambda$~Ori group, coloured by the difference between their abundance and the median of the group,
$\epsilon({\rm X}) - \tilde{\epsilon}({\rm X})$, in dex. Each panel has its own
colour scale, and the median subtracted is given in the colourbar label. The element
and, in parentheses, the number of stars with a determination are indicated in each
panel. Only stars not flagged as outliers are shown.}
    \label{fig:lori}
\end{figure*}

\textit{Orion A:} {We did not find any evidence of a significant correlation with RA or Dec in any of the elements (see Figure \ref{fig:a}). The strongest correlation is that of C with Dec ($r_s=-0.50$; $p_{\rm cor}=24\%$), while C and Fe show a weaker correlation with RA ($r_s=+0.40$ in both cases; $p_{\rm cor}=76\%$), so none of them is significant. Moran's $I$ gives an uncorrected $p$-value of 3.9\% for C, but once the Holm correction for the five tests is applied, such significance is lost  ($p_{\rm cor}=19.5\%$). Given the lower N abundance of this group (Sect.~\ref{sect:homo}), we paid special attention to this element. The N abundance shows no correlation with RA ($r_s=+0.13$; $p=60\%$) or Dec ($r_s=-0.18$; $p=45\%$), and the Moran's $I$ test finds no evidence of spatial structure ($p=32\%$), so the N deficit is not concentrated in any region of the group. As in the case of $\lambda$-Ori, our statistical tests are consistent with a chemically homogeneous group, although the number of stars is low (20 for C, N, O, and Fe, and 12 for S).}

\begin{figure*}
    \centering
    \includegraphics[width=\textwidth]{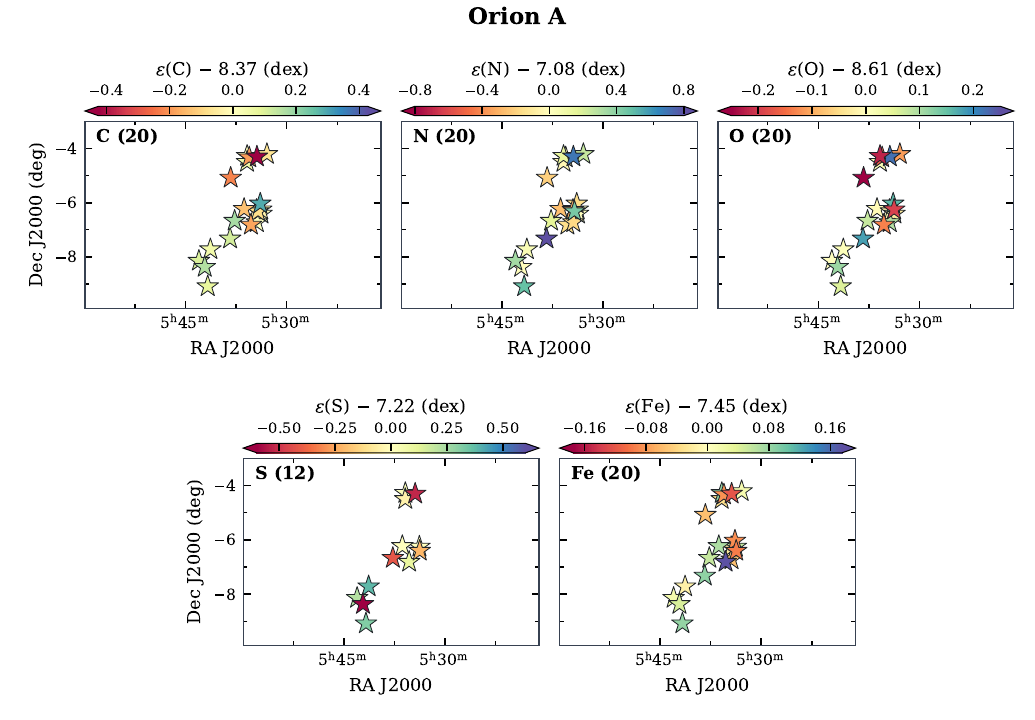}
    \caption{Same as Figure~\ref{fig:lori}, but for the Orion A group.}
    \label{fig:a}
\end{figure*}

\textit{Orion B:} {This group has the lowest number of stars: 17 for C, N, O, and Fe, and 9 for S. Similarly to Orion A, the moderate correlations found with RA, in this case for O ($r_s=+0.50$; $p=4.0\%$) and Fe ($r_s=-0.55$; $p=2.2\%$), did not survive the Holm correction: $p_{\rm cor}=36\%$ and $p_{\rm cor}=23\%$, respectively. The Moran tests did not find any spatial structure in the five elements. Fe is the element with the $p$-value (8.9\%) closest to the 5\% significance level we imposed, but after the Holm correction this value goes up to 45\%. As in the case of $\lambda$-Ori and Orion A, our results, as well as Figure \ref{fig:b}, show no evidence of chemical gradients in Orion B, although the small number of stars limits the sensitivity of the tests.  also}

\begin{figure*}
    \centering
    \includegraphics[width=\textwidth]{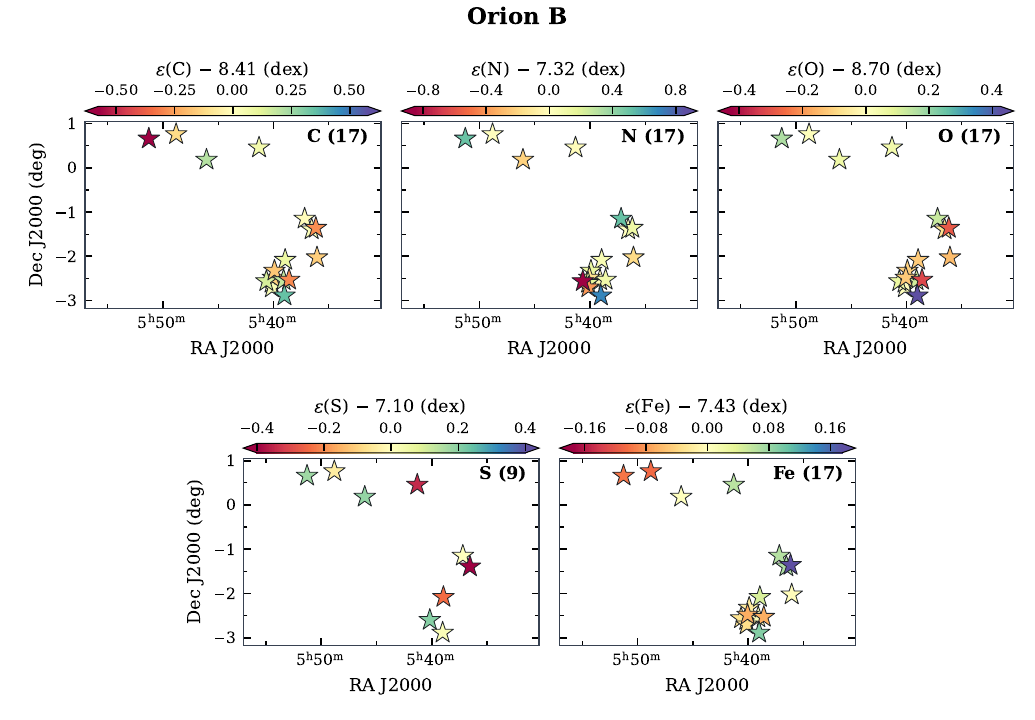}
    \caption{Same as Figure~\ref{fig:lori}, but for the Orion B group.}
    \label{fig:b}
\end{figure*}

\textit{Orion OB:} {This is the most populated group of our sample (118 stars for C, N, O, and Fe, and 88 for S), so the tests are more sensitive than for the other groups. We did not find any significant correlation between the chemical abundances and RA or Dec. The closest to the 5\% significance level is the correlation between RA and the S abundance ($r_s=+0.23$; $p=2.8\%$), but it loses its significance after the Holm correction ($p_{\rm cor}=28\%$). The Moran's $I$ tests do not show any evidence of spatial structure (see Figure \ref{fig:ob}) in the chemical abundances of Orion OB either, as in the rest of the groups.}

{In summary, the Spearman correlations of the abundances with RA and Dec and the Moran's $I$ tests do not reveal any significant spatial structure in the four groups of Orion. A few weak or moderate correlations appeared before applying the Holm correction, but none of them remains significant at the 5\% level once the number of tests performed in each group is taken into account ($p_{\rm cor} \geq 17\%$). The same holds for the Moran's $I$ tests, for which the lowest corrected $p$-value is 19.5\% (C in Orion A). In particular, the lower N abundance of Orion A does not show any spatial dependence within the group. These results are consistent with groups that are chemically homogeneous in C, N, O, S, and Fe, with no evidence of local gradients or clumps. 

However, given the small number of stars of Orion A (20), Orion B (17), and $\lambda$-Ori (40), these tests can only rule out large spatial variations, and the conclusion is more robust for Orion OB (118 stars). {In addition, our sample does not include stars in the innermost region of the ONC, and the precision of our abundances may not be sufficient to reveal subtle chemical differences, so we cannot exclude their presence, in particular in the immediate vicinity of the most massive stars.}

\begin{figure*}
    \centering
    \includegraphics[width=\textwidth]{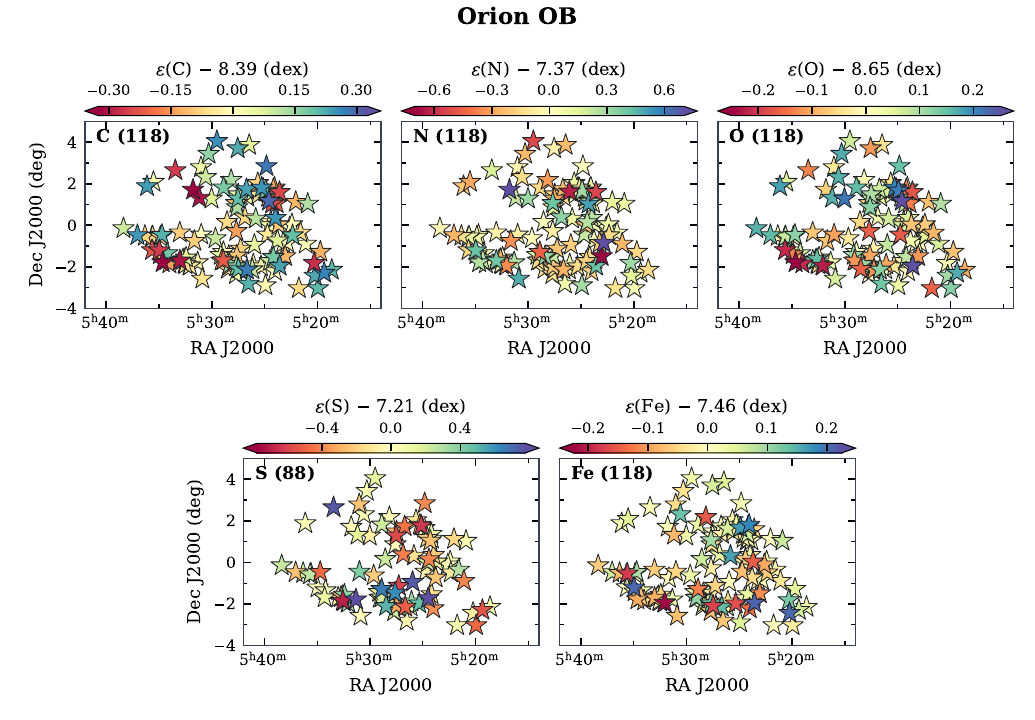}
    \caption{Same as Figure~\ref{fig:lori}, but for the Orion OB group.}
    \label{fig:ob}
\end{figure*}

\section{Comparison with the solar neighborhood and nebular regions}\label{sect:compar}
\subsection{Solar neighborhood}\label{sect:solar}
We used the main-sequence star sample of \cite{lopezval24} to establish the chemical abundances of the solar neighborhood, make comparisons with the Orion population of young stars, and establish the chemical differences between the two populations. \cite{lopezval24} derived the atmospheric parameters of over 1600 nearby (within 100~pc of the Sun) G-, K-, and M-type main-sequence stars (\teff\ between 3000 and 6000~K), and for most of them they also determined the abundances of Mg, Al, Si, Ca, and Fe. Using the stellar parameters reported in that work and the atomic and molecular lines listed in Table~\ref{tab:lines}, we derived the abundances of C, N, O, S, and Fe following the same procedure as for the young stars. {We kept for our analysis only those stars with C, N, O, and Fe determined from at least two spectral features. We were able to determine the abundances of C, N, O, and Fe in 430 main-sequence stars, while S was derived in 330 of those objects.}

{Before comparing the two samples, it is important to note that their atmospheric parameters are statistically different in the four cases (\teff, \logg, \meta, and \vsini; K-S $p_{\rm cor} < 0.001\%$), especially in \logg\ and \vsini, with the main-sequence stars having higher \logg\ and lower \vsini\ than the Orion stars. These differences are a direct consequence of the different evolutionary stages of the two samples: the young stars of Orion are still contracting toward the main sequence and therefore have lower surface gravities and, as is typical of such young objects, faster rotation than the main-sequence stars of the solar neighborhood. For this reason, the comparison presented below is qualitative: we compare the median abundances, with the median absolute deviation as the error, and not the full distributions. Note that the M-W and K-S tests detect small but statistically significant offsets between the two samples in all the elements ($p_{\rm cor} < 5\%$), but these offsets are smaller than the dispersion of each sample and can be expected given the different atmospheric parameters of the two populations. We divided the main-sequence sample into the same spectral-type bins used for Orion, obtaining 46 M stars (\teff\ $<$~4000~K), 286 K stars (4000~$\leq$~\teff~$\leq$~5000~K), and 98 G stars (\teff\ $>$~5000~K). The G-type stars are not included in the comparison, since the Orion sample contains practically no stars in that temperature range.}

\emph{Carbon} -- The median C abundance of the full main-sequence sample is slightly higher than the value found for the Orion stars, but still consistent within the errors. The median abundances of the M- and K-type main-sequence stars (Table~\ref{tab:mk_comparison}) are mutually consistent with each other, with the solar value, and with those measured in the M- and K-type stars of Orion, within the errors.

\emph{Nitrogen} -- The median N abundance of the full main-sequence sample is higher than the value measured for the Orion stars, but still consistent within the errors. When dividing the main-sequence sample into spectral types, the median abundances of the M- and K-type stars (Table~\ref{tab:mk_comparison}) are mutually consistent within the errors. {We also note from Table~\ref{tab:mk_comparison} that, while the C, O, S, and Fe abundances of the main-sequence stars are close to the solar values, N is clearly subsolar. The Orion stars show the same behavior, but the N depletion is smaller in the main-sequence sample. The [C/O] abundance ratios of the main-sequence stars ([C/O] = 0.02$\pm$0.06 dex) are close to zero and, together with the near-solar abundances of C, O, S, and Fe, suggest that the anomalous N abundance, likely a systematic negative offset, does not extend to the other abundances derived for the same stars.}

\emph{Oxygen} -- The median O abundance of the full main-sequence sample (Table~\ref{tab:mk_comparison}) is slightly higher than the value measured for the Orion stars, but still consistent within the errors. When separating the sample into spectral types, the median O abundances of the M- and K-type main-sequence stars are also consistent within the errors with those measured in the M- and K-type young stars of Orion.

\emph{Iron} -- The median Fe abundance derived for the whole main-sequence sample is in agreement with that measured for the Orion members. The median Fe abundance of the K-type stars is also consistent with that of the Orion stars, while the M-type stars show a slightly lower value, marginally consistent with Orion within the errors (see Table~\ref{tab:mk_comparison}).

\emph{Sulphur} -- The median S abundance of the {330} main-sequence stars with S determined (Table~\ref{tab:mk_comparison}) is in agreement with that measured in the Orion stars. We excluded the stars with \teff\ $<$~4000~K, as we did not consider the S abundances derived for these stars a good determination, so there are no M-type stars with S abundances in the sample. On the other hand, the median S abundance of the K-type main-sequence stars is consistent with that measured for the Orion K-type stars.

{In summary, despite the differences in the atmospheric parameters, the median abundances of C, N, O, S, and Fe are consistent within the errors between the young stars of Orion and the main-sequence sample (Table~\ref{tab:mk_comparison}), although C, N, and O are slightly higher in the main-sequence stars. The origin of this small offset is uncertain: they could reflect real differences in the composition of the gas from which the complex formed, but they could also arise from the different atmospheric parameters of the two samples, since the abundances of these elements are derived from molecular lines that are sensitive to \logg\ and \vsini\ (Table \ref{tab:param_dependence}). The fact that Fe and S, and the ratios [C/O] and [N/O], agree between both samples favors a common small offset in C, N, and O rather than a difference in their relative nucleosynthetic enrichment.} 

\subsection{Nebular regions}
We performed a first-order comparison between stellar and nebular abundances reported in the literature for a sample of ionized nebulae, with the aim of investigating the role of stellar nucleosynthesis in shaping the chemical composition of the molecular clouds from which star-forming regions originate.

\subsubsection{Dataset, physical conditions and chemical abundances}

The sample used in this work come from the DEep Spectra of Ionized REgions Database \citep[DESIRED;][]{desired2023}, which provides a statistically significant dataset based on deep optical spectra with direct determinatios of the electron temperature ($T_{\rm e})$ and includes HII regions, star forming galaxies, planetary nebulae, ring nebulae and Herbig-Haro objects. A detailed description of the spectral compilation, the homogenization of the analysis, the sample cleaning procedure, the main selection criteria, the complete list of detected gas emission lines, and the extinction correction methodology is reported by \cite{desired2023}. {Since spatially resolved nebular abundances in Orion at high spatial resolution are still a work in progress (Méndez-Delgado, in preparation), a subset of objects from the DESIRED sample is selected here as a reference for the chemical composition of ionized regions associated with young stars.} 

Based on the original DESIRED sample, the studies by \cite{mendezdelgadokreckelesteban2024} and \cite{estebanmendezdelgadogarcia2025} examined the Fe–N and Fe–O relations, and the S–O relation, respectively, restricting their analysis to HII regions in star-forming galaxies. Additionally, in the work by \cite{mendezdelgadokreckelesteban2024}, the data selection criteria required the detection of the [FeIII] $\lambda4658$ emission line. 

We adopt the nebular datasets presented in these two works to perform a comparison with our stellar abundance results, with the goal of improving our understanding of stellar feedback processes into the interstellar medium. The sample from \cite{mendezdelgadokreckelesteban2024} includes abundance determinations for 451 HII regions and star-forming galaxies, whereas the dataset from \cite{estebanmendezdelgadogarcia2025} comprises 3,154 HII regions and star-forming galaxies.

\begin{figure*}
    \centering
    \includegraphics[width=0.8\textwidth]{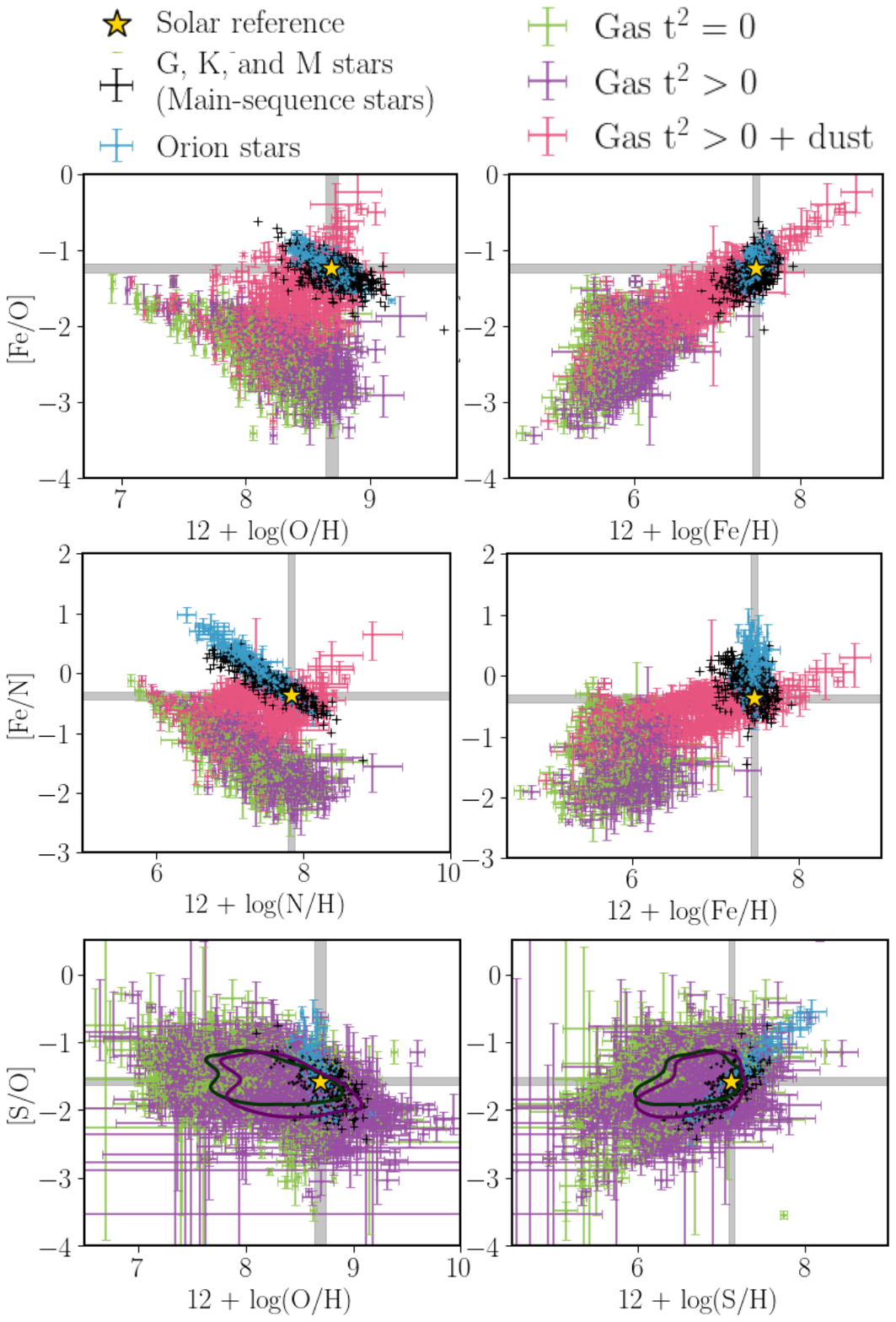}
    \caption{Comparison between stellar and nebular chemical abundances. The stellar abundances determined for the main-sequence sample and the Orion sample in this work are represented by black and blue stars, respectively. Gas-phase abundance estimates for the sample of HII regions and star forming galaxies from DESIRED are shown assuming $t^{2}=0$ (green stars), $t^{2}>0$ (purple stars), and in some cases $t^{2}>0$ including dust depletion (pink stars). In all panels, the solar abundances from \protect\cite{asplund21} are shown as a yellow star, with their corresponding uncertainties indicated by the gray shaded bands. {For visualization purposes, density contours were added to the lower panels to delineate regions enclosing approximately (68\%) of the corresponding nebular abundance distributions for the ($t^{2}=0$) (green contour) and ($t^{2}>0$) (purple contour) cases.}}
    \label{fig:gas_stars}
\end{figure*}

The determination of the physical conditions for the nebular samples was carried out using the \texttt{PyNeb} package (v1.1.18; \citealp{luridianamorissetshaw2015}), adopting a consistent set of atomic data together with the HI effective recombination coefficients from \cite{storeyhummer1995}. A detailed description of the application of the \texttt{getCrossTemDen}, \texttt{getTemDen}, and \texttt{getIonAbundance} routines, as well as the cross-consistency analysis of temperature-sensitive diagnostics, the Monte Carlo error propagation, and the criteria used to estimate a representative average electron density for each nebula, is provided in \cite{mendezdelgadokreckelesteban2024} and \cite{estebanmendezdelgadogarcia2025}.

{Both studies use the DESIRED direct-method framework, restricting the analysis to HII regions and star-forming galaxies with direct electron-temperature determinations and recalculating physical conditions and ionic abundances homogeneously across the samples \citep{mendezdelgadokreckelesteban2024, estebanmendezdelgadogarcia2025}. In the Fe/O–Fe/N work, unreliable electronic temperatures estimates were treated conservatively by privileging direct T$_{e}$([NII], 5755/6584) for the low-ionization zone because Fe$^{2+}$, O$^+$, and N$^+$ coexist there, and by warning that extrapolated T$_{e}$-T$_{e}$ relations become poorly tested at low metallicity and T$_{e} >$ 13000 K; when T$_{e}$([NII]) was unavailable, it was inferred from T$_{e}$([OIII], 4363/5007) and or T$_{e}$([SIII], 6312/9069) using the relations of \cite{garnett1992} for $t^{2} = 0$ and \cite{MendezDelgado23} for $t^{2} > 0$, propagating the same fractional uncertainty to the inferred temperature, while noting that this likely underestimates the true error and could shift Fe/O or Fe/N by up to $\sim$ 0.05 dex
\citep{mendezdelgadokreckelesteban2024}. Total O was obtained from the summed oxygen ionic abundances under the adopted physical conditions, whereas total N and Fe required ionization correction factors because not all relevant stages are observed directly. For S, the alpha-element \cite{estebanmendezdelgadogarcia2025} compared three ICF prescriptions: \cite{amayodelgadostasinska2021}, \cite{izotovstasinskameynet2006}, and \cite{dorsperezmonterohagele2016}, to select the most ionization-independent total abundances, with the schemes parameterized by (O$^{2+}$/O) and the \cite{izotovstasinskameynet2006} prescription applied only when (O$^{2+}$/O)$\sim$0.2. For N, the total abundance follows the standard low-ionization correction used in the DESIRED analyses, tying N to the O ionization structure after deriving N$^{+}$ in the same zone characterized by T$_{e}$([NII]). For Fe, the total abundance had to account for unobserved Fe$^{+}$ and Fe$^{3+}$; because direct [FeII] and [FeIV] lines are generally too weak or compromised, the preferred correction was the \cite{rodriguezrubin2005} ICF, which agreed better with direct multi-ion estimates than the \cite{izotovstasinskameynet2006} alternative, whose discrepancies could reach $\sim$ 0.7 dex, versus $\sim$ 0.2 dex for the work described by \cite{rodriguezrubin2005}, when $t^2 > 0$ is considered \citep{mendezdelgadokreckelesteban2024}. Finally, because Fe is strongly depleted into dust, the Fe study applied an additional dust-depletion correction by deriving a relation to estimate the fraction of Fe trapped in grains and adding that hidden component to the measured gas-phase Fe budget when interpreting total Fe/O and Fe/N trends \citep{mendezdelgadokreckelesteban2024}. This includes both the case of a homogeneous temperature structure ($t^{2}=0$) and an inhomogeneous one ($t^{2} > 0$; \citealp{Peimbert:1967}), as well as the $t^{2} > 0$ case including dust-depletion corrections derived for iron, given its strong tendency to be depleted onto dust grains (see \cite{mendezdelgadokreckelesteban2024}; their Equation 1)}.

Figure \ref{fig:gas_stars} shows the full comparison between stellar abundances in the local sample and Orion, and the nebular chemical abundances derived for the DESIRED subsamples. Throughout this section, the notation [X/Y] denotes $\epsilon$(X) - $\epsilon$(Y).


\subsubsection{Comparison for the case [Fe/O]}

The upper-left panel of Figure \ref{fig:gas_stars} corresponds to the determination of [Fe/O] as a function of 12+log(O/H) and shows that both stellar samples (main-sequence and Orion stars) follow a negative trend similar to that obtained from the nebular gas abundance determinations assuming $t^{2}=0$ and $t^{2}>0$ but with an offset of $\sim$1.5 dex in [Fe/O]. Instead, the stellar data points are broadly consistent with nebular abundance estimates that include dust depletion corrections but with a less defined trend. 

The observed negative trend in [Fe/O] appears to be primarily driven by a larger dispersion in oxygen abundances, while [Fe/H] remains comparatively homogeneous among the stellar samples. The behavior observed for the stars is consistent with previous analyses of Orion association, which reported little star-to-star variation in iron abundance but significant variations in oxygen abundance (e.g.  \citealp{cunhalambert1992, cunha98}). Such a pattern has been interpreted as evidence for self-enrichment within the Orion association, where ejecta from massive stars and Type II supernovae preferentially enhance oxygen while leaving the iron abundance largely unchanged \cite{cunhalambert1994, kos21}. In this context, the abundance trends observed in our young G-, K-, and M{-type} stars may provide additional support for localized chemical enrichment processes operating in the association, although we cannot rule out that part of the observed oxygen dispersion may also arise from the larger uncertainties affecting oxygen abundance estimates in low-mass stars.

On the other hand, in \cite{mendezdelgadokreckelesteban2024}, the comparison is performed using low-metallicity stellar samples, which naturally display a trend opposite to the one found in our analysis. In contrast, their Galactic B-type stars show a behavior more consistent with the trend observed here for our G-, K-, and M-type stellar samples. 

In the upper-right panel of Fig.~\ref{fig:gas_stars}, we present [Fe/O] as a function of 12+log(Fe/H). The nebular abundances show an increasing trend toward higher [Fe/H] values. In the case of the stellar samples, both the Orion and main sequence samples are consistent with the abundance estimates obtained for $t^{2}>0$ when dust depletion is taken into account, although both stellar samples span a more compact region of the plot.

The Orion G-, K-, and M-type stars are extremely young compared with the characteristic nucleosynthetic timescales responsible for oxygen and iron enrichment in the Galaxy. Oxygen is predominantly produced by core-collapse supernovae (CCSNe), originating from massive stars ($>$8 M$_{\odot}$) with lifetimes of a few tens of Myr ($\sim$50 Myr, \citealp{schneiderpodsiadlowskimuller2021}), whereas a significant fraction of Galactic iron is produced by Type Ia supernovae (SN Ia), which are associated with thermonuclear explosions of carbon–oxygen white dwarfs. Although the exact formation pathway of these events remains under debate, their progenitors evolve on delayed timescales relative to star formation, spanning a broad distribution with minimum delays of $\sim$40 Myr, corresponding to the time required for the formation of a white dwarf, and extending up to several Gyr \citep{maozmannucci2012}. Consequently, abundance trends driven by the progressive contribution of Type Ia supernovae are not expected to develop among the Orion members, given their young ages and small age dispersion. The observed [Fe/O] trend is therefore more likely to reflect either pre-existing chemical inhomogeneities in the parental gas, possibly associated with localized oxygen enrichment, and/or systematic effects related to the stellar abundance determinations. Instead, the behavior is more plausibly explained by stellar atmospheric effects influencing abundance derivations in cool pre–main-sequence stars. Strong magnetic activity, surface inhomogeneities such as starspots, chromospheric heating, and departures from LTE conditions can alter line formation, particularly for Fe I transitions, leading to systematic underestimation of iron abundances in cooler or more active stars \citep{bergemannlindcollet2012}. Furthermore, oxygen abundances derived from molecular OH lines are strongly sensitive to effective temperature, and the adopted atmospheric structure, particularly to three-dimensional temperature inhomogeneities and departures from LTE. Because OH formation depends on molecular equilibrium, small variations in the thermal structure of the atmosphere can translate into significant abundance shifts. These effects become especially relevant in young, magnetically active stars, where surface inhomogeneities, chromospheric heating, and enhanced activity may introduce systematic biases in the inferred oxygen abundances relative to those obtained from more robust indicators such as the forbidden [OI] line \citep{garciaperez06}. 

These combined effects can naturally generate correlated variations in [Fe/O] without requiring intrinsic chemical differences. Nevertheless, a secondary contribution cannot be entirely excluded: the parental gas of the Orion molecular cloud may have experienced prior local enrichment by massive stars and recent core-collapse supernovae, producing a modest oxygen enhancement in the material from which the current stellar population formed \citep{younggounellesmith2011}.

\subsubsection{Comparison for the case [Fe/N]}

The middle-left panel of Fig. \ref{fig:gas_stars} shows the determination of [Fe/N] as a function of the total nitrogen abundance, expressed as 12 + log(N/H). The negative trend displayed by the stellar abundance distributions is similar to that derived from the gas-phase abundances when no dust-depletion correction is applied ($t^{2}=0$ and $t^{2}>0$), although an offset is present between the stellar and nebular measurements. In terms of absolute abundance values, a better agreement is obtained when dust-depletion corrections are taken into account. The dust correction reduces the $\sim$1-dex shift in abundance but blurs the trend to the point of inverting its slope. Also the dust depletion corrected data for gas results in a larger range of [Fe/H] values as seen in the middle-right panel. When comparing the stellar abundances presented in this work (Orion and main sequence samples) with those reported by \cite{mendezdelgadokreckelesteban2024}, the agreement is stronger for the sample of B-type stars. {In contrast, the low-mass stellar sample exhibits a nearly flat behavior over the range 12+log([N/H]) $\sim$ 5-7, as shown in Fig. 8 of \cite{mendezdelgadokreckelesteban2024}, rather than the decreasing trend observed in the nebular abundances.}

Regarding the middle-right panel in the Fig. \ref{fig:gas_stars}, corresponding to [Fe/N] versus the total iron abundance expressed as 12 + log(Fe/H), both stellar abundance inferences again agree with the nebular abundance of N obtained for the case of $t^{2}>0$ including dust depletion. However, no clear trend or gradient is observed for the stars, while the dust-depletion corrected gas data points show some trend. The range of stellar abundances for Orion and the main-sequence sample (12 + log(Fe/H) $>$ 7 and an average value of [Fe/N] $\sim$ 0) is consistent with that reported by \cite{ecuvillonisraeliansantos2004}, which presents a sample of solar-type stars with and without confirmed planets. This behavior is consistent with the results of \cite{danielskibrucalassibenatti2022}, who report a mild increase; however, they do not confirm its statistical significance, as their measurements correspond to upper limits for stars in the 5000–6000 K range. Therefore, their study does not establish a conclusive trend, although it is not incompatible with the trend observed here.

However, the young G-, K-, and M-type stars in Orion are not expected to have significantly altered their surface nitrogen abundances through internal nucleosynthesis. The comparatively low nitrogen abundances relative to oxygen suggest that the chemical composition of the Orion stars may be influenced by enrichment from massive stars. Given the young age of the region, the delayed contribution of secondary nitrogen from intermediate-mass stars may not yet be fully reflected in the natal material from which these stars formed.

\subsubsection{Comparison for the case [S/O]}

The lower-left panel of Fig. \ref{fig:gas_stars} shows the [S/O] abundance as a function of the total oxygen abundance. To our knowledge, there is currently no published estimate of nebular abundances that includes a correction for dust depletion in this case and this is because it is assumed that the fraction of S depleted into dust is negligible in HII regions and star forming galaxies \citep{duffaucaffausbordonne2017}. The stellar abundances, particularly those corresponding to the main sequence sample, correlate with the nebular estimate assuming $t^{2}>0$. {The main-sequence stellar sample, the nebular sample ($t^{2}>0$), and the Orion stellar sample all exhibit negative trends. Since the relative strength of these correlations is not visually apparent in the comparison, we calculated Pearson's correlation coefficient r and the corresponding p-value for each sample to quantify the strength and statistical significance of the linear correlations. The resulting coefficients are r=-0.50 (p $<$ 0.01\%) for the main-sequence stellar sample, r=-0.38 (p $<$ 0.01\%) for the nebular sample, and r=-0.51 (p $<$ 0.01\%) for the Orion stellar sample. These results indicate statistically significant negative correlations in all three samples, with somewhat weaker linear correlation in the nebular sample than in the two stellar samples.}

The Orion stellar abundances derived in this work are consistent with the comparison presented by \cite{estebanmendezdelgadogarcia2025}, particularly with the Cepheid stars analyzed by \cite{dasilvadorazipalla2023}, whereas the B-type stellar abundances reported by \cite{wessmayerprzybillbutler2022} lie below the Orion distribution, reaching [S/O] values lower than -1.5. Also the negative trend is consistent with the work of \cite{diazzamora2022}. 

Finally, in the lower-right panel of Fig. \ref{fig:gas_stars} we show the [S/O] abundance as a function of the total sulphur abundance. We find a strong agreement general between the stellar samples (main sequence and Orion) and the nebular abundances assuming $t^{2} > 0$. A few stars in the Orion sample {(only the 18\%)} show elevated [S/O] values ([S/O] $>$ -1), while remaining consistent with the overall trend. Although [S/O] is not constant across the abundance range, the observed trends provide insight into the relationship between sulphur and oxygen abundances.
As in the case of [S/O] as a function of the total oxygen abundance, we calculated Pearson's correlation coefficient and the corresponding p-value to quantify the trends observed in the three samples. For the main-sequence stellar sample, the nebular sample ($t^{2}>0$), and the Orion stellar sample, we obtain r=0.75 (p $<$ 0.01\%), r=0.36 (p $<$ 0.01\%), and r=0.96 (p $<$ 0.01\%), respectively. These results indicate statistically significant positive correlations in all three samples, with the strongest linear correlation found for the Orion stellar sample and the weakest for the nebular sample. Since sulphur and oxygen are both predominantly produced by massive stars, their abundance patterns may provide complementary insights into the chemical enrichment history of the stellar populations.

{\subsubsection{Statistical Comparison of Stellar and Nebular Abundances}}

{We compared the stellar abundances in Orion with the nebular abundance datasets that showed the highest level of agreement with the stellar measurements. In particular, for O, Fe, and N, we considered the nebular abundance estimates that account for both dust depletion and temperature inhomogeneities ($t^{2}>0$), as these estimates span a parameter range that is most consistent with that covered by the stellar abundances. This provides a more suitable basis for a direct comparison between the stellar and nebular measurements. We used the M-W and K-S tests to assess differences in median values and overall distributions, respectively, on an element-by-element basis.}

{For oxygen and iron, the M-W test yields p $<$ 0.01\% for both elements. Nitrogen also exhibits a statistically significant difference between the stellar and nebular samples, although with a higher probability with p=1\%.}

{The K-S test reveals statistically significant differences between the stellar and nebular abundance distributions of oxygen, iron, and nitrogen, corresponding to p-values below 0.01\% for all three elements and D values of 0.57, 0.85, and 0.28, respectively. For oxygen and iron, the K-S results are consistent with the statistically significant differences detected by the M-W test. Nitrogen also exhibits statistically significant differences between the stellar and nebular abundance distributions according to both tests, although the M-W test yields a higher p-value (1\%) than the KS test (or below 0.01\%). This difference in statistical significance reflects the distinct aspects of the distributions assessed by the two tests: M-W compares the relative ranks of the observations, whereas KS evaluates differences between the empirical cumulative distribution functions. Overall, these results indicate that statistically significant differences between stellar and nebular abundances persist for all three elements, although their magnitude varies, as reflected in the KS statistics (D=0.57, 0.85, and 0.28 for oxygen, iron, and nitrogen, respectively).}

{Following the same statistical approach applied to iron, oxygen, and nitrogen, we completed the comparison between nebular and stellar abundances in Orion by analyzing sulphur. Since no sulphur abundance estimates corrected for dust depletion are currently available, this comparison was performed using nebular abundances that account only for temperature inhomogeneities ($t^{2}>0$). The M-W test yields U=288137.5 and p=0.01\%, indicating a highly significant difference between the stellar and nebular sulphur abundances. Consistently, the K-S test reveals a strong discrepancy between the overall distributions, yielding D=0.54 and below p=0.01\%. Together, these results indicate significant differences between the stellar and nebular sulphur abundance distributions, both in terms of location and overall shape.}

{Having established the overall statistical differences between the stellar and nebular abundance measurements by treating Orion as a single, integrated region, we next investigate the abundance distributions of its four stellar groups (A, B, OB, and $\lambda$Ori) individually, comparing each group with the nebular abundance measurements. For each element, we adopt the nebular abundance dataset that provides the best agreement with the Orion stellar abundances in the global comparison. To this end, we examine the differences between the median stellar abundances of each group and the median nebular abundances, as well as the results of the KS test, which assesses differences between the corresponding stellar and nebular abundance distributions. This complementary analysis allows us to investigate whether the discrepancies identified in the global comparison are shared by all four stellar groups or vary among them, thereby providing a more detailed assessment of the relationship between the stellar abundances of each group and those measured in the ionised gas.}

{To assess the robustness of the differences between the stellar and nebular oxygen abundances, we performed an independent bootstrap analysis of the stellar and nebular samples, generating 10,000 resampled datasets for each stellar group. We calculated the difference between the stellar and nebular median abundances in each realization and derived 95\% percentile confidence intervals from the resulting distributions.}

{Oxygen: The median stellar oxygen abundances are consistently higher than the median nebular abundance (12+log(O/H)=8.32) across all four groups, ranging from 8.606 dex in $\lambda$Ori to 8.702 dex in region B. The corresponding differences are 0.295, 0.382, 0.333, and 0.286 dex for regions A, B, OB, and $\lambda$Ori, respectively. The resulting of the bootstrap analysis in the 95\% confidence intervals for the differences between the stellar and nebular median abundances are [0.203, 0.362], [0.242, 0.453], [0.272, 0.372], and [0.213, 0.317] dex for regions A, B, OB, and $\lambda$Ori, respectively. All four intervals exclude zero, supporting the robustness of the positive median offsets to bootstrap resampling. The interval for region B is the broadest, consistent with its smaller stellar sample, whereas region OB yields a comparatively narrow interval. Overall, these results indicate that the discrepancy between stellar and nebular oxygen abundances persists across all four groups, although its magnitude and associated uncertainty vary among them. The KS test reveals statistically significant differences between the stellar and nebular abundance distributions for all four groups (p $<$ 0.01\%), with D values ranging from 0.537 to 0.644. Region OB exhibits the largest KS statistic (D=0.644), whereas $\lambda$Ori shows the smallest median offset (0.286 dex). These results indicate that the discrepancy between stellar and nebular oxygen abundances persists across all four groups, although its magnitude varies among them.}

{Iron: The differences between stellar and nebular iron abundances are substantially larger than those found for oxygen. The median stellar abundances range from 7.426 dex in region B to 7.462 dex in region OB, whereas the median nebular abundance is 6.660 dex. The corresponding offsets are 0.788, 0.766, 0.802, and 0.786 dex for regions A, B, OB, and $\lambda$Ori, respectively. The 95\% bootstrap confidence intervals for these differences are [0.694, 0.892], [0.672, 0.859], [0.720, 0.890], and [0.707, 0.879] dex, respectively. All four intervals exclude zero, supporting the robustness of the positive median offsets to bootstrap resampling. The KS test confirms highly significant differences between the stellar and nebular distributions in all four groups, with D values ranging from 0.849 to 0.882 and p $<$ 0.01\% in every case. Region OB exhibits the largest median offset (0.802 dex), whereas region B shows the smallest (0.766 dex). Overall, the consistently large median offsets, together with the bootstrap confidence intervals and the KS test results, indicate a pronounced discrepancy between stellar and nebular iron abundances across all four groups, with median offsets substantially larger than those found for oxygen.}

{Nitrogen: In contrast to oxygen and iron, nitrogen exhibits smaller differences between the median stellar and nebular abundances. The median nebular abundance is 7.205 dex, while the stellar medians range from 7.076 dex in region A to 7.371 dex in region OB. The corresponding absolute offsets are 0.129, 0.116, 0.166, and 0.154 dex for regions A, B, OB, and $\lambda$Ori, respectively. The 95\% bootstrap confidence intervals for these differences are [-0.241, 0.095], [-0.034, 0.258], [0.078, 0.241], and [0.032, 0.220] dex, respectively. The intervals for regions A and B include zero, indicating that their median offsets are not robustly distinguishable from zero under bootstrap resampling. In contrast, the intervals for regions OB and $\lambda$Ori are entirely positive, supporting higher stellar than nebular median nitrogen abundances in these two groups. The KS test does not reveal statistically significant differences for regions A (D=0.202, p=37.1\%) or B (D=0.307, p=7.3\%). However, statistically significant differences are found for region OB (D=0.342, p $<$ 0.01\%) and $\lambda$Ori (D=0.255, p=1.4\%). Overall, the bootstrap and KS results consistently support a stellar-nebular nitrogen abundance discrepancy in regions OB and $\lambda$Ori, while providing insufficient evidence for a statistically significant difference in regions A and B.}

{Sulphur: The median stellar sulphur abundances are systematically higher than the median nebular abundance of 6.671 dex, with stellar medians of 7.219, 7.103, 7.206, and 7.123 dex for regions A, B, OB, and $\lambda$Ori, respectively. The corresponding offsets are 0.548, 0.432, 0.536, and 0.452 dex. The 95\% bootstrap confidence intervals for these differences are [0.228, 0.705], [0.090, 0.619], [0.478, 0.601], and [0.356, 0.591] dex for regions A, B, OB, and $\lambda$Ori, respectively. All four intervals lie entirely above zero, supporting the robustness of the positive median offsets under bootstrap resampling. The uncertainty is comparatively large for regions A and B, which have the smallest stellar samples ($N=12$ and $N=9$, respectively), whereas region OB yields the narrowest confidence interval. Overall, these results indicate a systematic discrepancy between stellar and nebular sulphur abundances across all four groups, with positive median offsets ranging from 0.432 dex in region B to 0.548 dex in region A. The KS test also identifies statistically significant differences between the stellar and nebular abundance distributions in all four groups, with $D$ values ranging from 0.530 to 0.594 and p $<$ 0.8\% in every case. Region A exhibits both the largest median offset and the largest KS statistic, whereas region B shows the smallest values for both measures. Overall, these results indicate that the discrepancy between stellar and nebular sulphur abundances is present across all four groups, although its magnitude and uncertainty vary.}


\section{Summary}\label{sect:summary}
{We derived the abundances of C, N, O, S, and Fe for 226 young G-, K-, and M-type stars (\teff\ between 3500 and 5900~K) of the Orion complex, using APOGEE-2 spectra, MARCS model atmospheres, atmospheric parameters from \tona, and the \bacc\ code. Sulphur could be measured only in 173 of these stars. After removing 31 outliers, 195 slow rotators ($v\sin i\leq30$~km~s$^{-1}$) remained: 20 in Orion A, 17 in Orion B, 118 in Orion OB1, and 40 in $\lambda$~Ori. Because the S abundances of stars cooler than 4000~K are not reliable, we used only the 128 stars with \teff\ $\geq$ 4000~K for S. In all cases, we adopt the median as the representative value and the median absolute deviation as its uncertainty.}
{The median abundances of the whole sample are $\epsilon$(C) = 8.37$\pm$0.10, $\epsilon$(N) = 7.35$\pm$0.16, $\epsilon$(O) = 8.63$\pm$0.08, $\epsilon$(S) = 7.19$\pm$0.18, and $\epsilon$(Fe) = 7.45$\pm$0.05. Relative to the solar values of \citet{asplund21}, C, O, S, and Fe are within 0.09 dex of solar ([C/H] $=-0.09\pm0.11$, [O/H] $=-0.06\pm0.09$, [S/H] $=0.07\pm0.18$, and [Fe/H] $=-0.01\pm0.06$). In contrast, N is clearly subsolar, with [N/H] $=-0.48\pm0.17$.}

{Within the Orion groups, we found no significant correlation between the abundances and RA or Dec, and the Moran's $I$ tests do not indicate spatial structure (the lowest corrected $p$-value is 17\%). The groups span an age range of about 8 Myr, from 3.2~Myr in Orion B to 11.4~Myr in Orion OB1. The Kruskal-Wallis tests find no difference between the groups for C, S, and Fe ($p$ = 19, 35, and 45\%), whereas N and O show lower $p$-values ($p$ $\sim$ 3\% and 0.06\%, respectively). After applying the Holm correction to the pairwise tests, the only significant difference involving Orion A is a lower N abundance, which is 0.29 dex below that of Orion OB1 ($p_{\rm cor}$ = 1.2\%) and corresponds to [N/H] $=-0.75\pm0.18$, compared with values between $-0.51\pm0.17$ and $-0.46\pm0.17$ in the other groups. The median C, O, S, and Fe abundances of Orion A differ from those of Orion OB1 by only $-0.02$, $-0.04$, 0.01, and $-0.01$ dex. Moreover, the differences in \teff\ ($\sim$96~K), \logg\ ($\sim$0.13 dex), [M/H] ($\sim$0.01 dex), and $v\sin i$ ($\sim$0.1~km~s$^{-1}$) between the two groups would only produce a difference of $\sim$0.04 dex in N. The O abundances of Orion OB1 and $\lambda$~Ori are also different ($p_{\rm cor}$ = 0.74\%), but the difference is only $\sim$0.05 dex and is compatible with the $\sim$0.03 dex expected from the lower \teff\ ($\sim$200~K) of $\lambda$~Ori.}

{We compared Orion with 430 nearby main-sequence stars (330 with S) analysed in the same way (46 M-type and 286 K-type stars, excluding the G-type stars, which are not represented in Orion). Their median abundances are $\epsilon$(C) = 8.48$\pm$0.13, $\epsilon$(N) = 7.56$\pm$0.23, $\epsilon$(O) = 8.73$\pm$0.12, $\epsilon$(S) = 7.12$\pm$0.14, and $\epsilon$(Fe) = 7.43$\pm$0.10, with [N/H] $=-0.27\pm0.24$. Compared with Orion, C, N, and O are higher in the main-sequence stars by 0.11, 0.21, and 0.10 dex, whereas S and Fe differ by $-0.07$ and $-0.02$ dex, respectively. The [C/O] ratios of the two samples are also consistent ($-0.03\pm0.06$ in Orion and $0.02\pm0.06$ in the main-sequence stars). All these differences are smaller than the dispersion of each sample. Although the Mann-Whitney and Kolmogorov-Smirnov tests detect statistically significant offsets ($p_{\rm cor}<5\%$), the atmospheric parameters of the two samples also differ significantly (Kolmogorov-Smirnov $p_{\rm cor}<0.001\%$), mostly in \logg\ and $v\sin i$, so we only consider this comparison qualitative.}

{We also compared the stellar abundances with those of H {\sc ii} regions and star-forming galaxies from the DESIRED database. The stellar abundances are systematically higher than the nebular ones assuming $t^2>0$, with median offsets of 0.29--0.38 dex for O, 0.43--0.55 dex for S, and 0.77--0.80 dex for Fe, and with the smallest offset for N (0.12--0.17 dex). The stellar and nebular [Fe/O] and [Fe/N] ratios agree better when dust depletion is included, but the stellar and nebular distributions are statistically different for O, Fe, N, and S (K-S $D$ = 0.57, 0.85, 0.28, and 0.54, with $p<0.01\%$ in all cases). The [S/O] ratios show similar trends, although the M-W and K-S tests also detect significant differences ($p<0.01\%$).}

\section{Conclusions}\label{sect:conclusions}
{Using APOGEE-2 spectra, we derived the C, N, O, S, and Fe abundances of 195 young stars in four groups of the Orion complex: Orion A (20 stars), Orion B (17), Orion OB1 (118), and $\lambda$~Ori (40). We could not measure significant chemical differences between these groups beyond our uncertainties, which is consistent with them having formed from chemically homogeneous gas, given their short age spread ($\sim$8 Myr) compared with the timescales of interstellar medium enrichment. Because of the small number of stars in Orion A, Orion B, and $\lambda$~Ori, our tests can only rule out large spatial variations, and subtle differences may exist below our detection limits.}

{The subsolar N abundance is the main exception to the solar composition of the complex. It also appears in 430 nearby main-sequence stars analysed in the same way and does not extend to C, O, S, or Fe, so it is likely a systematic negative offset in the N abundances derived from the molecular lines, and not a nucleosynthetic signature, since stars of this age (below 11.4 Myr) are not expected to have altered their surface N through internal nucleosynthesis. The N abundance of Orion A is lower than that of the other groups, and this difference cannot be explained by the differences in atmospheric parameters or by an age sequence, so a larger sample in this group is required to determine its origin.}

{In the comparison between the whole Orion sample and the main-sequence stars, the offsets of 0.1--0.2 dex in C, N, and O could reflect real differences or arise from the different atmospheric parameters of the two samples. However, the agreement between the two samples in Fe, S, and [C/O] favours a small shift affecting C, N, and O together, rather than a difference in their relative nucleosynthetic enrichment. Finally, the agreement between the stellar abundances and those of HII regions and star-forming galaxies improves when temperature inhomogeneities and dust depletion are considered in the nebular abundances, although the stellar and nebular distributions remain statistically different. This highlights the need to account simultaneously for temperature inhomogeneities, dust depletion, and stellar atmospheric effects when comparing young stars with ionised gas. Resolved nebular abundances in Orion and similar analyses in other young regions are natural next steps to test these points.}

\section*{Acknowledgements}
We thank the anonymous referee for their helpful comments and suggestions. R. L-V. and L. A. acknowledge support from Secretar\'ia de Ciencia, Humanidades, Tecnolog\'ia e Inovacci\'on (SECIHTI) through a postdoctoral fellowship within the program ``Estancias posdoctorales por M\'exico''. AZL-A gratefully acknowledges the support provided by the Postdoctoral Program (POSDOC) of UNAM (Universidad Nacional Autónoma de México). JEMD, AZL-A and L. C. thank the support by UNAM/DGAPA/PAPIIT/IA103326 project “DESIRED (DEep Spectra of Ionized Regions Database): de las emisiones más sutiles a la física fundamental del universo”. C. R-Z. acknowledges support from UNAM-DGAPA-PAPIIT grant IN 107226. J.G.F-T and R.R.L. gratefully acknowledge the support provided by ANID Fondecyt Regular No. 1260371. S.V. gratefully acknowledges the support provided by the ANID BASAL project FB210003. Funding for the Sloan Digital Sky Survey IV has been provided by the Alfred P. Sloan Foundation, the U.S. Department of Energy Office of Science, and the Participating Institutions. SDSS-IV acknowledges support and resources from the Center for High-Performance Computing at the University of Utah. The SDSS website is www.sdss.org. 
SDSS-IV is managed by the Astrophysical Research Consortium for the Participating Institutions of the SDSS Collaboration including the Brazilian Participation Group, the Carnegie Institution for Science, Carnegie Mellon University, the Chilean Participation Group, the French Participation Group, Harvard-Smithsonian Center for Astrophysics, Instituto de Astrof\'isica de Canarias, The Johns Hopkins University, Kavli Institute for the Physics and Mathematics of the Universe (IPMU) / University of Tokyo, Lawrence Berkeley National Laboratory, Leibniz Institut f\"ur Astrophysik Potsdam (AIP), Max-Planck-Institut f\"ur Astronomie (MPIA Heidelberg), Max-Planck-Institut f\"ur Astrophysik (MPA Garching), Max-Planck-Institut f\"ur Extraterrestrische Physik (MPE), National Astronomical Observatories of China, New Mexico State University, New York University, University of Notre Dame, Observat\'ario Nacional / MCTI, The Ohio State University, Pennsylvania State University, Shanghai Astronomical Observatory, United Kingdom Participation Group, Universidad Nacional Aut\'onoma de M\'exico, University of Arizona, University of Colorado Boulder, University of Oxford, University of Portsmouth, University of Utah, University of Virginia, University of Washington, University of Wisconsin, Vanderbilt University, and Yale University.  
This research has made use of the SIMBAD database, operated at CDS, Strasbourg, France.

This work made use of the high-throughput computing (HTC) infrastructure and the optimisation of computing resources of the project Grid UNAM (\url{https://grid.unam.mx/}) at the Universidad Nacional Autónoma de México. Participating Institutions of the project Grid UNAM includes Dirección General de Cómputo y de Tecnologías de Información y Comunicación (DGTIC), Laboratorio de Modelos y Datos Científicos (LAMOD), Instituto de Ciencias de la Atmósfera y Cambio Climático (ICAyCC), Instituto de Astronomía (IA) and Instituto de Ciencias Nucleares (ICN). We also thank the members of technical and development staff of the project Grid UNAM for their continuous support.

\section*{Data Availability}
Table \ref{tab:res} is available in the online supplementary material of the paper and at CDS via anonymous ftp to cdsarc.u-strasbg.fr (130.79.128.5) or via https://cdsarc.cds.unistra.fr/viz-bin/cat/J/MNRAS/Vol/num.



\bibliographystyle{mnras}
\bibliography{biblio} 




\appendix
\section{Outlier identification}\label{app:out}
{To identify outlier abundance determinations, we computed a modified $z$-score for each star, for each element, and for each region as
\begin{equation}
M_i = 0.6745\,\frac{x_i - \tilde{x}}{\rm MAD},
\end{equation}
where $x_i$ is the abundance of the analysed element in the $i$-th star, and $\tilde{x}$ and MAD are the median and the median absolute deviation of the abundances of that element in the whole population of the selected region. We flagged as candidate outliers the stars with $|M_i| > 3.5$, the threshold recommended by \citet{iglewicz93}, which affected 12 of the 1077 abundance measurements (1.1\%), in 10 of the 226 stars. }

{Since a statistical outlier is not necessarily a bad measurement, we compared each flagged star by its abundance uncertainty, number of lines used, \teff, and \vsini\ to decide whether the deviation was more likely an artifact or a real abundance.}

{Six of the ten flagged stars lie at the edges of our temperature range. Only four stars have $T_{\rm eff}>5400$~K in our sample, and three of them, located in the Orion OB group, are flagged: 2M05270085-0117575 (\teff\ = 5718 K), 2M05292524-0030133 (\teff\ = 5749 K), and 2M05271018+0050452 (\teff\ = 5477 K), in Fe, N, and O, respectively. The fourth, 2M05390088-0253163 (Orion B), is not flagged, but has C, N, and O abundances 0.5--0.7 dex above its group median ($M_i=2.0$--$3.2$). At these temperatures (\teff $\gtrsim$ 5400 K), the molecular lines used in our analysis start to weaken. At the cool edge, 2M05341389+0923394 ($\lambda$~Ori; \teff\ = 3334~K, the coolest of the sample), 2M05453826-0008109 (Orion B; \teff\ = 3646~K), and 2M05251672+0116158 (Orion OB; \teff\ = 3440~K) are flagged as outlier. In the first two, the outlier abundace value comes from a single line with an uncertainty four to five times the typical one, 2M05341389+0923394 also has all its elements above the median. 2M05251672+0116158 is low in C, O, and Fe by 0.5--0.7~dex, and also in N, consistent with its fitted [M/H]$=-0.38$ dex; however, a visual inspection reveals a noisy spectrum in most of the spectral regions used.}

{Three more stars exceed the threshold only marginally ($|M|=3.6$): 2M05343548-0427211 (in C), 2M05325664+0018478 (in Fe), and 2M05225965+0120089 (in S). The first two rotate at 25.7 and 24.8 km s$^{-1}$, well above the median of the sample (19.0 km s$^{-1}$) and close to our upper limit of 30 km s$^{-1}$, which may make their lines harder to measure; in 2M05325664+0018478, the Monte Carlo distributions of all five Fe lines are also broad and skewed ($\sigma=0.11$--$0.45$~dex, compared with a typical 0.02--0.03~dex). 2M05225965+0120089 is the most ambiguous. S is the most difficult element to measure in our sample, since only two S lines were selected in our study, and it shows the largest dispersion of the five elements (MAD$=0.22$~dex in OB). In this star, only one of the two S lines is usable, so the S abundance rests on a single measurement. However, visual inspection shows a good fit to the line and no obvious artifacts in the spectrum. Moreover, its abundance uncertainty, temperature, and rotation are normal.}

{Finally, 2M05253368+1138489 (located in $\lambda$~Ori) shows no sign of a measurement problem in the Fe lines. Its temperature (4256 K) and rotation (15.6 km s$^{-1}$) are typical, and its $\epsilon$(Fe)$=7.15$, 0.29~dex below the group median, comes from five lines with a small uncertainty. Its other elements are also slightly below the median, consistent with its low global metallicity ([M/H]$=-0.34$ dex). Although this star could be genuinely metal-poor relative to its group, we flagged it as an outlier.}

{We exclude these outliers from further analysis, together with the stars whose C, N, O, or Fe abundance comes from a single spectral line, since such determinations have no internal consistency check. Single-line stars are over-represented among the outliers (40\% of them, compared with 11\% of the sample, i.e., 25 of 226 stars) and are responsible for three of the cases discussed above. This criterion preferentially removes cool stars: 21 of the 25 have $T_{\rm eff}\leq 4000$~K. }

\bsp	
\label{lastpage}
\end{document}